\documentclass[%
 reprint,
superscriptaddress,
reprint,
 amsmath,amssymb,
 aps,
 prl,
floatfix,
]{revtex4-2}

\usepackage{graphicx}
\usepackage{dcolumn}
\usepackage{bm}
\usepackage{hyperref}

\DeclareMathAlphabet{\mathpzc}{OT1}{pzc}{m}{it}

\DeclareMathAlphabet{\mathpzc}{OT1}{pzc}{m}{it}

\newcommand{\p}{\partial}
\newcommand{\1}{\frac{1}{2}}

\newcommand{\cD} {\mathcal{D}}

\newcommand{\cG} {\mathcal{G}}

\newcommand{\sqg}{\sqrt{g}}
\newcommand{\bp}{\bar{\partial}}

\newcommand{\cL} {\mathcal{L}}

\usepackage[dvipsnames]{xcolor}
\usepackage{soul,ulem}
\usepackage{upgreek}

\begin{document}


\title{Geometry fluctuations in chiral superfluids}

\author{Gabriel Cardoso}
\email{gabriel.cardoso@sjtu.edu.cn}
\affiliation{Tsung-Dao Lee Institute,
Shanghai Jiao Tong University, Shanghai, 201210, China}
\author{Qing-Dong Jiang}
\affiliation{Tsung-Dao Lee Institute,
Shanghai Jiao Tong University, Shanghai, 201210, China}
\affiliation{ School of Physics and Astronomy, Shanghai Jiao Tong University, Shanghai 200240, China}
\affiliation{Shanghai Branch, Hefei National Laboratory, Shanghai 201315, China}

\date{\today}

\begin{abstract} 
The coupling of chiral superfluids and superconductors to the background geometry leads to surprising geometric induction phenomena. We show that this coupling bears important consequences even in a nearly flat background, through its signature in thermal fluctuations. Starting from the Ginzburg-Landau free energy of a chiral superfluid minimally coupled to the background geometry, we show that the interaction strength between vortices gets renormalized by geometry fluctuations. In our setup, these arise from the shape fluctuations of the underlying two-dimensional substrate, and are controlled by its bending rigidity and tension. In the tensionless limit, the fluctuations lower the vortex interaction strength at large distances, which leads to a lowering of the BKT transition temperature. We study this effect in terms of the renormalization group flow of a dual sine-Gordon theory of the superfluid transition coupled to the substrate shape. It shows that, in turn, the chiral superfluid order can suppress the amplitude of shape fluctuations, resulting in an extended phase diagram which links the superfluid transition to the crumpling transition of the substrate. These reveal thermodynamic signatures of chirality in the superfluid transition.

\end{abstract}

\maketitle

\begin{figure}[h]
    \centering 
    \includegraphics[width=0.45\textwidth]{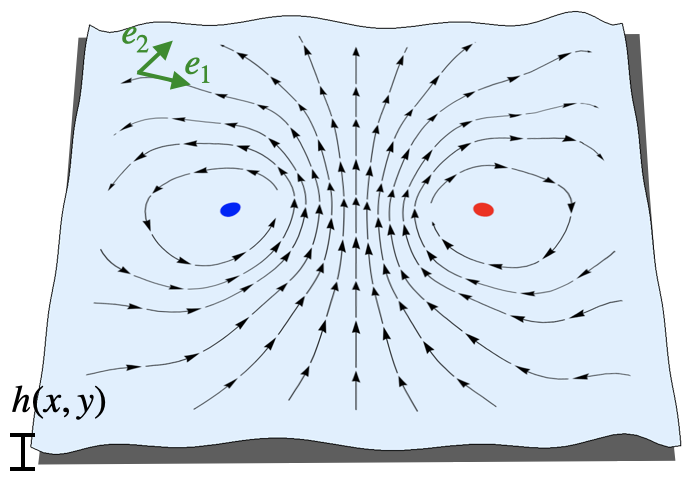}
    \caption{Illustrated is a chiral superfluid on a flexible two-dimensional substrate, with the gradient of the phase field (black) plotted around a vortex-antivortex pair (blue and red). The height of the membrane is parameterized by $h(x,y)$. Geometric fluctuations on a flat background renormalize the confining interaction between the vortex and antivortex.}
    \label{fig:vortexgeometry}
\end{figure}

\noindent\textit{Introduction} --Chiral phases of matter display a rich interplay between topology, geometry, and anomalies \cite{shapere1989geometric,wen1992shift,qi2010chiral,gromov2015framing,xiong2015evidence,nissinen2020emergent}. A particular case of interest is that of chiral superfluids and superconductors, whose topological defects carry anyonic zero modes with interesting braiding properties \cite{read2000paired,volovik2000monopoles,sato2016majorana,qi2011topological,tsutsumi2008majorana}. In these phases, the chiral order parameter couples to the background geometry \cite{moroz2015effective,kvorning2018proposed,jiang2020geometric}, and the resulting geometric induction effects can provide an avenue for manipulating topological defects \cite{qingdong}. 
However, little attention has been given to the role of geometry fluctuations. As explored in this letter, averaging over thermal fluctuations of the background geometry can lower the strength of the vortex-antivortex confining potential, modifying the phase diagram of chiral superfluidity. Our setup is illustrated in figure \ref{fig:vortexgeometry}. 
While we assume the two-dimensional substrate on which the superfluid condensate forms to be nearly flat, its material only has finite rigidity and tension, indicating inevitable shape fluctuations at nonzero temperatures.

On the other hand, the amplitude of shape fluctuations is governed by the renormalized properties of the substrate, such as bending rigidity, tension, and elastic coefficients. These properties determine the phase diagrams of various classes of thermalized surfaces \cite{bowick2001statistical,nelson2004statistical,helfrich1973elastic,1982degennes,nelson1987fluctuations,guitter1990tethering,shankar2021thermalized,metayer2022three}, and specifically, the critical properties of the low-temperature flat phase that have recently been experimentally investigated \cite{nicholl2017hidden,lopez2022effect}.
We find that the presence of chiral superfluidity modifies the renormalization group flow of substrate properties, typically strengthening the bending rigidity. Our analysis is based on the Ginzburg-Landau theory of a chiral superfluid condensate minimally coupled to the background geometry, allowing us to derive the energetic characteristics of vortex defects. 
We then analyse the dual sine-Gordon theory describing the condensation of these defects and the resulting renormalization group flow coupled to the surface degrees of freedom. 
While we focus on chiral superfluids, experimentally realized in the Helium-3 A-phase, our model serves as a prototype for understanding geometric and elastic effects in chiral superconductors and, more broadly, chiral phases.

\vspace{.4cm}
\noindent\textit{Vortex-geometry interaction} --
On a curved surface, the Ginzburg-Landau free energy for a superfluid condensate is given by
\begin{equation}
    \int dt d^2x\sqrt{g}\left[\Psi^\dagger \left(iD_t+\frac{1}{2m}D_iD^i\right)\Psi-V(|\Psi|)\right],\label{eq:minaction}
\end{equation}
\begin{figure}[h!]
    \centering    \includegraphics[width=0.45\textwidth]{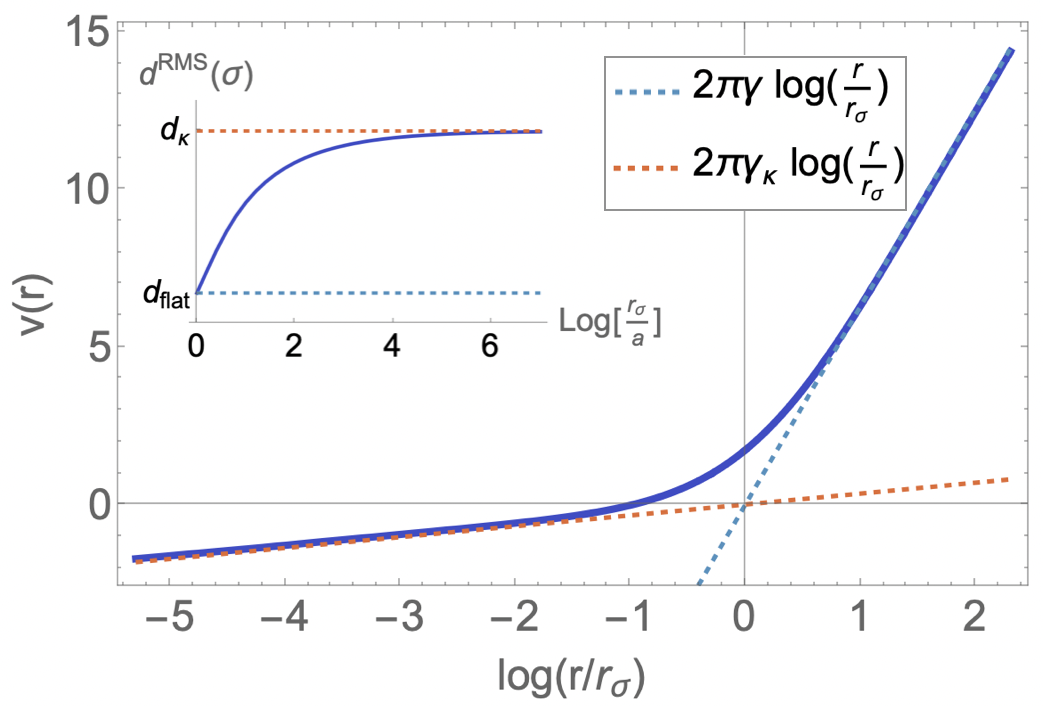}
    \caption{Modified interaction potential in a vortex-antivortex pair. The interaction strength $\gamma_\kappa$ is renormalized at scales smaller than $r_\sigma\sim\sqrt{\kappa/\sigma}$ relative to the flat background coefficient $\gamma$. Inset: average vortex-antivortex distance in confined pairs as a function of $r_\sigma/a$, showing the increase of the mean distance in the low tension limit.}
    \label{fig:fullpotential}
\end{figure}
where $D_\mu$ stands for the covariant derivative. Using a local basis for the tangent space $e_{1,2}$ (figure \ref{fig:vortexgeometry}), the chiral order parameter $\Psi$ can be expressed as a complex wavefunction, $\Psi=\sqrt{\rho}e^{i\theta}\epsilon_{+}$, where $\epsilon_{\pm}=\frac{1}{2}(e_1\pm i e_2)$ and we choose the positive chirality configuration without loss of generality. 
This parametrizes the chiral superfluid in terms of a scalar function, yet the coupling to geometry remains through the covariant derivative 
$D_\mu(\psi \epsilon_{+})=(\p_\mu+ \ell\omega_\mu)\psi \epsilon_{+}$, where $\omega_\mu=e_1\cdot\p_\mu e_2$ is the spin connection and $\ell$ is the condensate angular momentum in units of $\hbar$ (set to one). In the superfluid phase, the potential $V(\rho)\sim \alpha(\rho-\bar{\rho})^2$ gives an expectation value to the condensate density $\bar{\rho}>0$. Integrating out the density fluctuations yields
\begin{equation}
    \cL = \frac{\gamma_0}{2}(\p_0\theta+\ell\omega_0)^2+\frac{\gamma}{2}(\p_i\theta+\ell\omega_i)^2,\label{eq:lagrangian}
\end{equation}
describing the fluctuations of the phase field $\theta$, with $\gamma_0=\hbar^2/2\alpha$ and the effective superfluid stiffness $\gamma=\hbar^2\bar{\rho}/m$. In the following, we focus on the $p$-wave case where $\ell=1$.

We consider the system in the canonical ensemble, with the partition function given by:
\begin{equation}
    Z=\sum_n e^{-\beta E_n}=\int\cD\theta\cD g e^{-\frac{\tilde{\gamma}}{2}\int d^2x\sqg(\p_i\theta+\omega_i)^2}e^{-\beta E_g},
\end{equation}
where $\beta E_g$ represents the Euclidean action for the geometric degrees of freedom, and  $\tilde{\gamma}=\beta\gamma$ denotes the ratio with the thermal energy $k_BT$. Let us first consider the superfluid phase part by writing $Z=\int\cD g Z_\theta e^{-\beta E_g}$. The $2\pi$-periodicity of the $\theta$ field allows for vortex configurations - localized regions around which $\theta$ winds by a multiple of $2\pi$,   
$\oint_\Gamma d\theta = 2\pi q, \,\, q\in\mathbb{Z}.\label{eq:vcirculation}$
A vortex ( $q=1$) and an antivortex ($q=-1$) are shown in figure \ref{fig:vortexgeometry}. To integrate over the configurations of the $\theta$ field, we first introduce a Hubbard-Stratonovich dual field $\xi^j$,
\begin{equation}
    Z_\theta=\int\cD\theta\cD\xi e^{-\int d^2x\sqg\left[\frac{\xi^2}{2\tilde{\gamma}}+i\xi^j(\p_j\theta+\omega_j)\right]}.
\end{equation}
Next, we decompose the field $\theta=\theta_v+\theta_s$ into the multi-valued vortex component $\theta_v$ and the single-valued fluctuations $\theta_s$. Integrating out the smooth part $\theta_s$ imposes the constraint $D_j\xi^j=0$, which can be solved by introducing a scalar potential $\phi$ through $\xi^j=\epsilon^{jk}D_k \phi$. Consequently, the partition function becomes
\begin{equation}
    Z_\theta=\int\cD\theta_v\cD \phi e^{-\int d^2x\sqg\left[\frac{(\p\phi)^2}{2\tilde{\gamma}}+i\phi(\rho_v+\rho_g)\right]},\label{eq:thetavintegral}
\end{equation}
where we introduced the vortex and geometric charges,
\begin{align}
    &\rho_v=\epsilon^{ij}\p_i\p_j\theta_v, &\rho_g=\epsilon^{ij}\p_i\omega_j.\label{eq:rhog}
\end{align}
From the definition of the spin connection, it follows that
$\rho_g$ is given by the Gaussian curvature of the background $R$ \cite{supplementary}. Note that for a configuration of vortices with circulation $q_n=\pm 1,\pm 2,...$ at positions $r_n$, $\rho_v=\frac{1}{\sqg}\sum_{i}2\pi q_i\delta(r-r_i)$, and integrating out the field $\phi$ in (\ref{eq:thetavintegral}) results in the Boltzmann factor $e^{-\beta E_C}$, where
\begin{align}
    E_C = \sum_{i<j}e^2 q_iq_j v_g(r_i,r_j)+\sum_i e^2 q_i W_g(r_i),\label{eq:vortexgas}
\end{align}
with $v_g$ being the surface Coulomb potential (ie. the Green's function of the Laplace-Beltrami operator on the background metric $\Delta_g$), $W_g$ represents the electrostatic potential generated by the charge distribution $\rho_g=R$, and the elementary charge $e$ is given by $2\pi\sqrt{\gamma}$. Thus for a chiral superfluid on a curved background the mapping to the two-dimensional Coulomb plasma of vortices \cite{chuilee1975} is modified in a two different ways: through the modification of the vortex interaction potential $v_g$ and through the effective background charge generated by the curvature $R$. In particular, the screening of the curvature charge by vortices provides a geometric mechanism for manipulating vortex configurations \cite{qingdong}. We note that these follow quite generally from the minimal coupling of the order parameter to the background geometry, and that in principle the vortex cores can also couple non-minimally to the geometry \cite{turner2010vortices,wiegmann2014anomalous}.

\vspace{.4cm}
\noindent\textit{Renormalized vortex potential} -- Vortices play an important role in the superfluid phase transition. On a flat background, the potential $v(x,y)$ increases logarithmically with distance, making free vortices thermodynamically unfavored below the Berezinskii-Kosterlitz-Thouless (BKT) transition temperature \cite{KT}. However, equation (\ref{eq:vortexgas}) shows that a flexible substrate can compensate a vortex charge with an opposite geometric charge, by bending into the third direction. Interestingly, even for a very rigid, nearly flat substrate, this coupling can still lower the energy cost of a vortex when averaged over geometry fluctuations. To estimate this effect, we consider the effective potential $\Gamma[\varphi;\rho_v] = W[\eta;\rho_v]-\int d^2x\varphi\eta$, which is the Legendre transform of the generating functional $W[\eta;\rho_v]=-\log
    \langle e^{\int d^2x\phi\eta} \rangle$ evaluated at fixed vortex distribution $\rho_v$.
The effective potential $\Gamma[\varphi,\rho_v]$, also known as the quantum effective action, has the property that its saddle point in $\varphi=\langle\phi\rangle$ includes the full effect of field fluctuations.

\begin{figure}[h!]
    \centering    \includegraphics[width=0.45\textwidth]{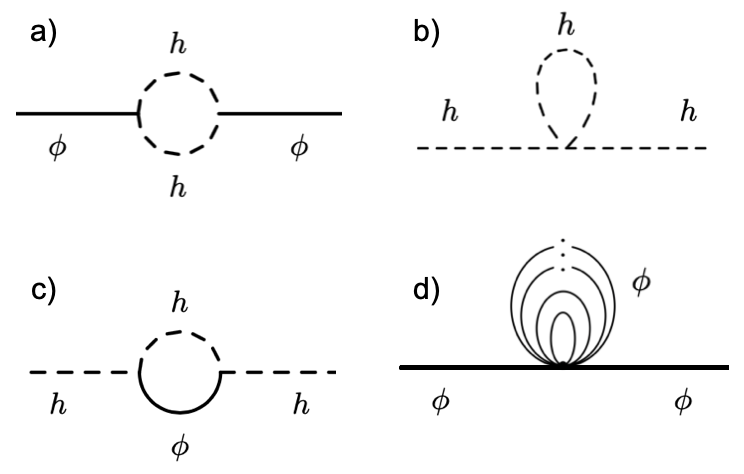}
    \caption{Some of the one-loop diagrams coupling the chiral superfluid to the background geometry. a) renormalization of $\phi$ propagator by height fluctuations; b) higher-order corrections to the bending energy lower the bending rigidity; c) conversely, the presence of chiral superfluid order can increase the bending rigidity at large distances; d) tadpole diagrams stemming from lowest order in the vortex fugacity expansion.}
    \label{fig:diagrams}
\end{figure}

The average over background geometries is evaluated with respect to the energy cost of shape configurations of the two-dimensional substrate, $E_g$. In terms of the embedding map coordinates, $\vec{r}(x,y)$, the metric induced on the surface of the substrate is $g_{ij}=\p_i \vec{r}\cdot\p_j \vec{r}$, and the leading contributions to the energy are expressed in terms of local geometric quantities as \cite{helfrich1973elastic}
\begin{equation}
    E_g=\sigma\int d^2x\sqrt{g}+\kappa\int d^2x\sqrt{g}\frac{H^2}{2}+\kappa_G\int d^2x\sqrt{g}R,\label{eq:bendingenergy}
\end{equation}
where $H$ is the mean curvature, defined as the trace of the local curvature tensor \cite{supplementary}. The isotropic tension $\sigma$ corresponds to the energy cost of changing the surface area, the bending rigidity $\kappa$ to the cost of generating extrinsic curvature and the Gaussian rigidity $\kappa_G$ to the cost of generating intrinsic curvature. For small deformations of a nearly flat membrane, one can use the Monge parametrization $\vec{r}(x,y)=(x,y,h(x,y))$ in terms of a smoothly modulated height field $h(x,y)$ (see figure \ref{fig:vortexgeometry}). In this case, the topological $\kappa_G$ term is irrelevant, and the other terms can be expanded in the gradients of $h$,
\begin{equation}
    E_g=\frac{1}{2}\int d^2x[\kappa(\Delta h)^2+\sigma(\p h)^2].\label{eq:Egh}
\end{equation}
To leading order in height fluctuations, we then find
\begin{align}
    &\Gamma[\varphi;\rho_v]=\int\left[-\frac{1}{2}\varphi\left(\frac{\Delta}{\tilde{\gamma}}-\langle R_hR_h\rangle_h\right)\varphi+i\varphi\rho_v\right],\label{eq:gammapotential}
\end{align}
where $R_h=\frac{1}{2}[(\Delta h)^2-(\p_{ij}^2 h)^2]$
is the small-gradient approximation to the Gaussian curvature. The second contribution corresponds to the 1-loop diagram in figure \ref{fig:diagrams} a). In the limit of small tension $\sigma\to 0$, one finds $\langle R_h(q) R_h(-q)\rangle_h=\frac{3q^2}{16(2\pi)^3}\tilde{\kappa}^{-2}$, which in the effective potential (\ref{eq:gammapotential}) corresponds to a renormalization of the superfluid stiffness $\gamma\to\gamma_\kappa$, with
\begin{equation}
    \gamma_\kappa^{-1} = \gamma^{-1} + \frac{3T}{32\pi\kappa^2}.\label{eq:gammashift}
\end{equation}
For a vortex pair configuration $\rho_v=2\pi \delta(r-r_1)-2\pi \delta(r-r_2)$, minimizing the effective potential $\Gamma[\varphi;\rho_v]$ over $\varphi$ gives
\begin{align}
    \Gamma = 2E_c+2\pi\gamma_\kappa\log\Big|\frac{r_1-r_2}{a}\Big|,
\end{align}
where we reguralized the singular part of the energy by introducing the vortex core size $a$ and the core energy $E_c$. As in the case of a flat background, the energy of a vortex-antivortex pair increases logarithmically with separation, which can lead to their confinement in the low-temperature superfluid phase through the BKT transition mechanism. The effect of geometry fluctuations is to reduce the prefactor $\gamma_\kappa$, which gives the energy scale of the vortex-antivortex interaction. As seen from equation (\ref{eq:gammashift}), this effect is stronger at higher temperatures and for softer membranes (smaller $\kappa$), for which thermal fluctuations of the background are more intense. In the case of finite membrane tension $\sigma>0$, we see from (\ref{eq:Egh}) that the amplitude of shape fluctuations is highly suppressed on length scales beyond $r_\sigma=\sqrt{\kappa/\sigma}$. As a result, we find that the vortex interaction potential is only renormalized for $r\lesssim r_\sigma$. The full potential is shown in figure \ref{fig:fullpotential}. An observable signature of the modified pontential is the dependence of the average size of vortex-antivortex pairs $d^{\rm RMS}=\sqrt{\langle(r_1-r_2)^2\rangle}$ on the background tension, which we plot in the inset. We see that it strongly depends on the tension, increasing from the flat-background result $d_{\rm flat}=a\sqrt{1-(2-\pi\tilde{\gamma})^{-1}}$ at large tension $r_\sigma\to a$ to the renormalized value $d_\kappa=a\sqrt{1-(2-\pi\tilde{\gamma}_\kappa)^{-1}}$ at small tension $r_\sigma\gg a$ \cite{supplementary}. While $d_{\rm flat}$ diverges at the usual BKT transition value $T_{\rm BKT}=\frac{\pi}{2}\gamma$, $d_\kappa$ diverges at a lower temperature determined by the renormalized superfluid stiffness $\gamma_\kappa$. This indicates that the softening of the potential by the background fluctuations can facilitate the vortex proliferation, leading to a lower transition temperature.

\vspace{.4cm}
\noindent\textit{Sine-Gordon map and RG flow} -- In the low-tension limit, we found that the vortex-antivortex confining potential is softened by the substrate shape fluctuations, which can modify the critical temperature for vortex unbinding. To understand the modified phase diagram, we consider the renormalization group flow of the coupled degrees of freedom in terms of a dual sine-Gordon theory. We evaluate the integral over the vortex part of the field $\theta_v$ in (\ref{eq:thetavintegral}) as a sum over vortex configurations $\rho_v=\frac{1}{\sqg}\sum_{i}2\pi q_i\delta(r-r_i)$,
\begin{equation}
    Z_\theta=\int\cD\phi e^{-\int d^2x\sqg\left[\frac{(\p\phi)^2}{2\tilde{\gamma}}+i R\phi\right]}\prod_{r_n}\sum_{q_n}e^{-i 2\pi q_n\phi(r_n)-\tilde{E}_c q_n^2}.\label{eq:zphi}
\end{equation}
The configurations contributing to this sum are the ones corresponding to the neutral vortex plasma $0=\int d^2x\sqg (\rho_v+\rho_g)=2\pi(Q+\chi)$, where the Euler characteristic $\chi$ appears through the Gauss-Bonnet theorem, and $Q=\sum_i q_i$ is the total charge of the vortex configuration. Performing the sum (\ref{eq:zphi}) over such vortex distributions leads to the modified sine-Gordon action
\begin{align}
    S_\phi=\int d^2x\sqg\left[\frac{(\p\phi)^2}{2\tilde{\gamma}}+i R\phi-\frac{2z}{a^2}\cos(2\pi\phi)\right],\label{eq:sinegordon}
\end{align}
where we defined the vortex fugacity $z=e^{-\tilde{E}_C}$.  For simplicity, we took the torus case $\chi=0$, but the more general case is discussed in the supplementary material \cite{supplementary}. Note that the coupling term is consistent with the discrete shift symmetry in $\phi$ due to the integer nature of the Euler characteristic \cite{riemannsurface}.

We consider the effective potential $\Gamma$ evaluated from the full partition function
\begin{equation}
    \int \cD\phi\cD h e^{-\int d^2x\left[\frac{(\p\phi)^2}{2\tilde{\gamma}}+i R_h\phi-\frac{2z}{a^2}\cos(2\pi\phi)+\frac{\tilde{\kappa}(\Delta h)^2}{2}+V_h\right]},
\end{equation}
as an expansion in the vortex fugacity $z$, in the deviation of $\tilde{\gamma}$ from the free boson fixed point $\tilde{\gamma}=2/\pi$, and in the inverse bending rigidity $\kappa^{-1}$.  Here, $V_h$ denotes corrections to the energy of the substrate. For example, the next-to-leading-order terms of the small-gradient expansion of the bending energy (\ref{eq:bendingenergy}) give $V_h=\tilde{\kappa}\left[2\Delta h \bar{\p}^i h\p^2_{ij}h\bar{\p}^j h-\frac{5}{2}(\Delta h)^2(\nabla h)^2\right]$, where $\bar{\p}^i=\epsilon^{ij}\p_j$. At one loop, this vertex leads to the diagram in figure \ref{fig:diagrams} b), which lowers the value of the bending rigidity $\kappa$ at large distances. This reflects the fact that in free membranes, for which the energy is given exclusively by the bending energy, the bending rigidity vanishes at large distances, so that the height field $h(x,y)$ has a finite correlation length. The membrane is said to be crumpled by the strong shape fluctuations. The presence of chiral superfluid order can counterbalance the renormalization of the bending rigidity, which at one loop corresponds to the diagram in figure \ref{fig:diagrams} c). Finally, the expansion in vortex fugacity corresponds to the characteristic tadpole diagrams \cite{coleman75sinegordon} such as in figure \ref{fig:diagrams} d). Regularization of the effective potential $\Gamma$ leads to the one-loop renormalization group flow equations \cite{supplementary}
\begin{figure}[h!]
\label{fig:superfluidflow}
    \centering \includegraphics[width=0.45\textwidth]{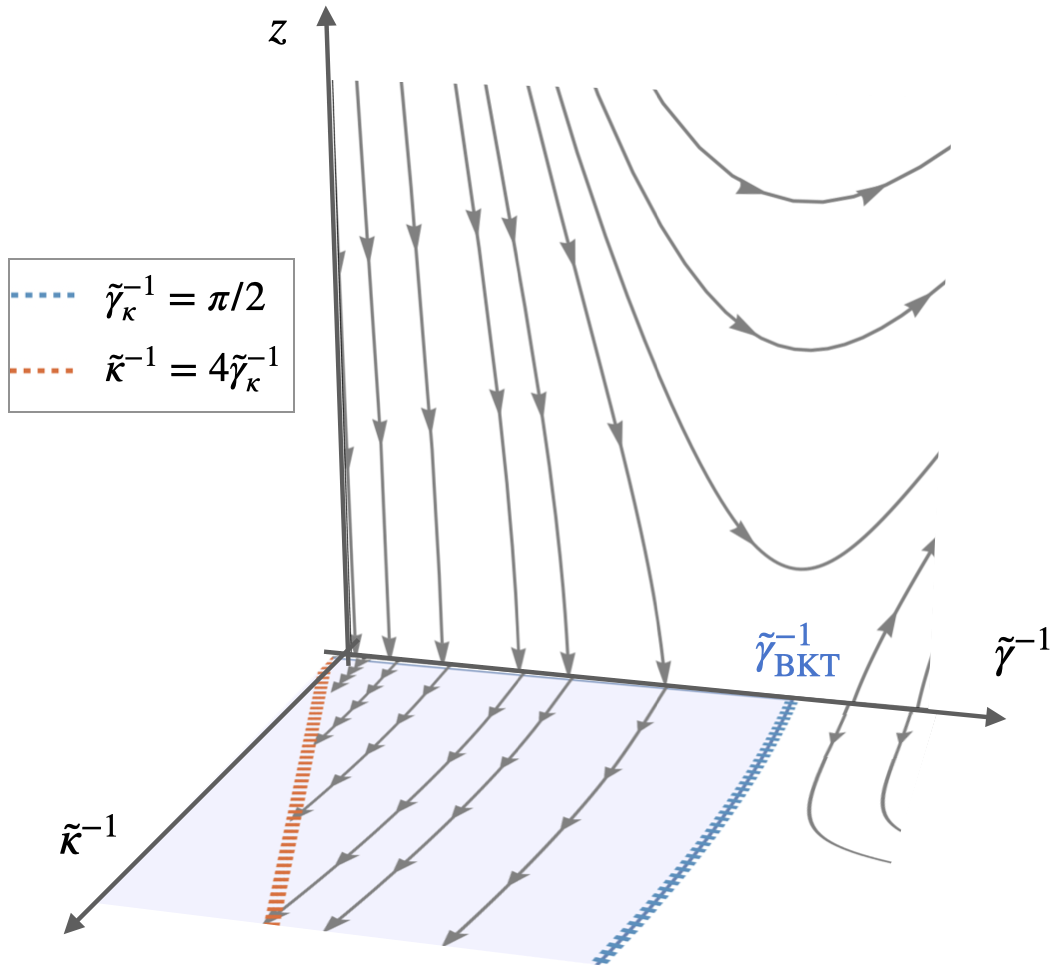}
    \caption{Renormalization group flow in the $(\tilde{\gamma}^{-1},z,\tilde{\kappa}^{-1})$ space. The flat limit $\tilde{\kappa}^{-1}\to 0$ corresponds to the flow diagram of the BKT transition. At finite bending rigidity $\kappa^{-1}>0$, the BKT transition point extends to the line $\tilde{\gamma}_\kappa=\frac{2}{\pi}$, indicating a lowering of the transition temperature. Below the transition, the system flows to the fixed line $\tilde{\kappa}^{-1}=4\tilde{\gamma}^{-1}$.}
    \label{fig:flowdiagram}
\end{figure}
\begin{align}
    \frac{d\tilde{\gamma}_{\kappa}^{-1}}{dl}&=4\pi^3 z^2,\label{eq:BKT1}\\
    \frac{dz}{dl}&=\left(2-\pi\tilde{\gamma}_{\kappa}\right)z.\label{eq:BKT2}\\
    \frac{d\tilde{\kappa}}{dl}&=-\frac{3}{4\pi}\left(1-\frac{\tilde{\gamma}_{\kappa}}{4\tilde{\kappa}}\right).\label{eq:kappaflow}
\end{align}
The resulting RG flow trajectories in the $(\tilde{\gamma}^{-1},z,\tilde{\kappa}^{-1})$ space are shown in figure \ref{fig:flowdiagram}. Equations (\ref{eq:BKT1},\ref{eq:BKT2}) differ from the flat background case by the renormalization $\gamma\to\gamma_\kappa$, which leads to a $\kappa$-dependent shift in the BKT transition temperature,
\begin{equation}
    \frac{\Delta T}{T_{\rm BKT}^0}=-\frac{3\gamma^2}{64\kappa^2}.
\end{equation}
On the other hand, equation (\ref{eq:kappaflow}) shows that the renormalization group flow of the bending rigidity $\kappa$ is modified by the presence of chiral superfluid order: instead of flowing to zero, the bending rigidity flows to the fixed line $\kappa=\gamma_\kappa/4$. This fixed line has an analogy in the phase diagram of hexatic membranes, where the quasi-long-range orientational order stabilizes the membrane fluctuations \cite{david1987critical,nelson1987fluctuations,KTpaticsurface,park1996sine}. In our case, the superfluid correlations stabilize the substrate shape fluctuations, leading to a flat phase below the BKT transition.

The presence of crystalline or hexatic in-plane order in the substrate leads to additional contributions to the vertex $V_h$. These can suppress shape fluctuations and lead to a flat phase even in the absence of the chiral superfluid \cite{bowick2001statistical}. At one loop (corresponding to lowest order in $\kappa^{-1}$), we find that the extra in-plane order modifies equation (\ref{eq:kappaflow}), but not (\ref{eq:BKT1},\ref{eq:BKT2}), which gives a good setup in which to study the modification of the flat phase properties by the chiral superfluid transition. In the case of a hexatic substrate, we find that the formation of the superfluid enhances the flat phase, corresponding to a positive shift in the value of the bending rigidity at large distances \cite{supplementary}. This shows that there are elasto-geometric signatures of the formation of a chiral condensate. Similarly, the case of a crystalline background is an interesting question for future investigation.

\vspace{.4cm}
\noindent\textit{Discussion} -- We examined the statistical mechanical consequences of the coupling to background geometry in chiral superfluids. The vortex-antivortex interaction strength gets renormalized below the length scale $r_\sigma\sim \sqrt{\kappa/\sigma}$, which separates the bending-dominated from the tension-dominated regime of the substrate shape fluctuations. The softening of the potential leads to an increase in the mean vortex-antivortex separation, which can trigger the BKT transition in the tensionless limit. We studied this question in terms of a dual sine-Gordon theory of the vortex-unbinding transition, which leads to a coupling of the chiral superfluid phase diagram to that of the substrate geometry, in particular to the crumpling transition. The result is a lowering of the BKT transition temperature parametrized by the bending rigidity and, conversely, an enhancement of the flat phase of the background parametrized by the superfluid density. These thermodynamic properties are elasto-geometric signatures specific to the formation of a chiral superfluid, distinguishing it from the scalar case of $s$-wave condensates. The low tension regime of strong shape fluctuations has been experimentally observed in suspended two-dimensional materials such as graphene \cite{nicholl2017hidden,lopez2022effect}, which makes them the right platform on which to test our ideas. Moreover, they display anomalous elastic coefficients \cite{david1988crumpling,aronovitz1988fluctuations,le1992self,mauri2020scaling}, which we expect to be modified upon formation of the chiral condensate, similarly to our result on the case of a hexatic background. We leave this interesting question for future work.

\vspace{.4cm}
\noindent\textit{Acknowledgements}--We are grateful to Alexander Abanov, Vadim Grinenko and Sudhakantha Girmohanta for useful discussions. This work was supported by National Natural Science Foundation of China (NSFC) under Grant No. 23Z031504628 (G.C. and Q.-D.J.), Jiaoda2030 Program Grant No.WH510363001, TDLI starting up grant, and Innovation Program for Quantum Science and Technology Grant No.2021ZD0301900 (Q.-D.J.).

\bibliography{apssamp}

\begin{thebibliography}{41}%
\makeatletter
\providecommand \@ifxundefined [1]{%
 \@ifx{#1\undefined}
}%
\providecommand \@ifnum [1]{%
 \ifnum #1\expandafter \@firstoftwo
 \else \expandafter \@secondoftwo
 \fi
}%
\providecommand \@ifx [1]{%
 \ifx #1\expandafter \@firstoftwo
 \else \expandafter \@secondoftwo
 \fi
}%
\providecommand \natexlab [1]{#1}%
\providecommand \enquote  [1]{``#1''}%
\providecommand \bibnamefont  [1]{#1}%
\providecommand \bibfnamefont [1]{#1}%
\providecommand \citenamefont [1]{#1}%
\providecommand \href@noop [0]{\@secondoftwo}%
\providecommand \href [0]{\begingroup \@sanitize@url \@href}%
\providecommand \@href[1]{\@@startlink{#1}\@@href}%
\providecommand \@@href[1]{\endgroup#1\@@endlink}%
\providecommand \@sanitize@url [0]{\catcode `\\12\catcode `\$12\catcode `\&12\catcode `\#12\catcode `\^12\catcode `\_12\catcode `\%12\relax}%
\providecommand \@@startlink[1]{}%
\providecommand \@@endlink[0]{}%
\providecommand \url  [0]{\begingroup\@sanitize@url \@url }%
\providecommand \@url [1]{\endgroup\@href {#1}{\urlprefix }}%
\providecommand \urlprefix  [0]{URL }%
\providecommand \Eprint [0]{\href }%
\providecommand \doibase [0]{https://doi.org/}%
\providecommand \selectlanguage [0]{\@gobble}%
\providecommand \bibinfo  [0]{\@secondoftwo}%
\providecommand \bibfield  [0]{\@secondoftwo}%
\providecommand \translation [1]{[#1]}%
\providecommand \BibitemOpen [0]{}%
\providecommand \bibitemStop [0]{}%
\providecommand \bibitemNoStop [0]{.\EOS\space}%
\providecommand \EOS [0]{\spacefactor3000\relax}%
\providecommand \BibitemShut  [1]{\csname bibitem#1\endcsname}%
\let\auto@bib@innerbib\@empty
\bibitem [{\citenamefont {Shapere}\ and\ \citenamefont {Wilczek}(1989)}]{shapere1989geometric}%
  \BibitemOpen
  \bibfield  {author} {\bibinfo {author} {\bibfnamefont {A.}~\bibnamefont {Shapere}}\ and\ \bibinfo {author} {\bibfnamefont {F.}~\bibnamefont {Wilczek}},\ }\href@noop {} {\emph {\bibinfo {title} {Geometric phases in physics}}},\ Vol.~\bibinfo {volume} {5}\ (\bibinfo  {publisher} {World scientific},\ \bibinfo {year} {1989})\BibitemShut {NoStop}%
\bibitem [{\citenamefont {Wen}\ and\ \citenamefont {Zee}(1992)}]{wen1992shift}%
  \BibitemOpen
  \bibfield  {author} {\bibinfo {author} {\bibfnamefont {X.-g.}\ \bibnamefont {Wen}}\ and\ \bibinfo {author} {\bibfnamefont {A.}~\bibnamefont {Zee}},\ }\bibfield  {title} {\bibinfo {title} {Shift and spin vector: New topological quantum numbers for the hall fluids},\ }\href@noop {} {\bibfield  {journal} {\bibinfo  {journal} {Physical review letters}\ }\textbf {\bibinfo {volume} {69}},\ \bibinfo {pages} {953} (\bibinfo {year} {1992})}\BibitemShut {NoStop}%
\bibitem [{\citenamefont {Qi}\ \emph {et~al.}(2010)\citenamefont {Qi}, \citenamefont {Hughes},\ and\ \citenamefont {Zhang}}]{qi2010chiral}%
  \BibitemOpen
  \bibfield  {author} {\bibinfo {author} {\bibfnamefont {X.-L.}\ \bibnamefont {Qi}}, \bibinfo {author} {\bibfnamefont {T.~L.}\ \bibnamefont {Hughes}},\ and\ \bibinfo {author} {\bibfnamefont {S.-C.}\ \bibnamefont {Zhang}},\ }\bibfield  {title} {\bibinfo {title} {Chiral topological superconductor from the quantum hall state},\ }\href@noop {} {\bibfield  {journal} {\bibinfo  {journal} {Physical Review B}\ }\textbf {\bibinfo {volume} {82}},\ \bibinfo {pages} {184516} (\bibinfo {year} {2010})}\BibitemShut {NoStop}%
\bibitem [{\citenamefont {Gromov}\ \emph {et~al.}(2015)\citenamefont {Gromov}, \citenamefont {Cho}, \citenamefont {You}, \citenamefont {Abanov},\ and\ \citenamefont {Fradkin}}]{gromov2015framing}%
  \BibitemOpen
  \bibfield  {author} {\bibinfo {author} {\bibfnamefont {A.}~\bibnamefont {Gromov}}, \bibinfo {author} {\bibfnamefont {G.~Y.}\ \bibnamefont {Cho}}, \bibinfo {author} {\bibfnamefont {Y.}~\bibnamefont {You}}, \bibinfo {author} {\bibfnamefont {A.~G.}\ \bibnamefont {Abanov}},\ and\ \bibinfo {author} {\bibfnamefont {E.}~\bibnamefont {Fradkin}},\ }\bibfield  {title} {\bibinfo {title} {Framing anomaly in the effective theory of the fractional quantum hall effect},\ }\href@noop {} {\bibfield  {journal} {\bibinfo  {journal} {Physical review letters}\ }\textbf {\bibinfo {volume} {114}},\ \bibinfo {pages} {016805} (\bibinfo {year} {2015})}\BibitemShut {NoStop}%
\bibitem [{\citenamefont {Xiong}\ \emph {et~al.}(2015)\citenamefont {Xiong}, \citenamefont {Kushwaha}, \citenamefont {Liang}, \citenamefont {Krizan}, \citenamefont {Hirschberger}, \citenamefont {Wang}, \citenamefont {Cava},\ and\ \citenamefont {Ong}}]{xiong2015evidence}%
  \BibitemOpen
  \bibfield  {author} {\bibinfo {author} {\bibfnamefont {J.}~\bibnamefont {Xiong}}, \bibinfo {author} {\bibfnamefont {S.~K.}\ \bibnamefont {Kushwaha}}, \bibinfo {author} {\bibfnamefont {T.}~\bibnamefont {Liang}}, \bibinfo {author} {\bibfnamefont {J.~W.}\ \bibnamefont {Krizan}}, \bibinfo {author} {\bibfnamefont {M.}~\bibnamefont {Hirschberger}}, \bibinfo {author} {\bibfnamefont {W.}~\bibnamefont {Wang}}, \bibinfo {author} {\bibfnamefont {R.~J.}\ \bibnamefont {Cava}},\ and\ \bibinfo {author} {\bibfnamefont {N.~P.}\ \bibnamefont {Ong}},\ }\bibfield  {title} {\bibinfo {title} {Evidence for the chiral anomaly in the dirac semimetal na3bi},\ }\href@noop {} {\bibfield  {journal} {\bibinfo  {journal} {Science}\ }\textbf {\bibinfo {volume} {350}},\ \bibinfo {pages} {413} (\bibinfo {year} {2015})}\BibitemShut {NoStop}%
\bibitem [{\citenamefont {Nissinen}(2020)}]{nissinen2020emergent}%
  \BibitemOpen
  \bibfield  {author} {\bibinfo {author} {\bibfnamefont {J.}~\bibnamefont {Nissinen}},\ }\bibfield  {title} {\bibinfo {title} {Emergent spacetime and gravitational nieh-yan anomaly in chiral p+ i p weyl superfluids and superconductors},\ }\href@noop {} {\bibfield  {journal} {\bibinfo  {journal} {Physical review letters}\ }\textbf {\bibinfo {volume} {124}},\ \bibinfo {pages} {117002} (\bibinfo {year} {2020})}\BibitemShut {NoStop}%
\bibitem [{\citenamefont {Read}\ and\ \citenamefont {Green}(2000)}]{read2000paired}%
  \BibitemOpen
  \bibfield  {author} {\bibinfo {author} {\bibfnamefont {N.}~\bibnamefont {Read}}\ and\ \bibinfo {author} {\bibfnamefont {D.}~\bibnamefont {Green}},\ }\bibfield  {title} {\bibinfo {title} {Paired states of fermions in two dimensions with breaking of parity and time-reversal symmetries and the fractional quantum hall effect},\ }\href@noop {} {\bibfield  {journal} {\bibinfo  {journal} {Physical Review B}\ }\textbf {\bibinfo {volume} {61}},\ \bibinfo {pages} {10267} (\bibinfo {year} {2000})}\BibitemShut {NoStop}%
\bibitem [{\citenamefont {Volovik}(2000)}]{volovik2000monopoles}%
  \BibitemOpen
  \bibfield  {author} {\bibinfo {author} {\bibfnamefont {G.}~\bibnamefont {Volovik}},\ }\bibfield  {title} {\bibinfo {title} {Monopoles and fractional vortices in chiral superconductors},\ }\href@noop {} {\bibfield  {journal} {\bibinfo  {journal} {Proceedings of the National Academy of Sciences}\ }\textbf {\bibinfo {volume} {97}},\ \bibinfo {pages} {2431} (\bibinfo {year} {2000})}\BibitemShut {NoStop}%
\bibitem [{\citenamefont {Sato}\ and\ \citenamefont {Fujimoto}(2016)}]{sato2016majorana}%
  \BibitemOpen
  \bibfield  {author} {\bibinfo {author} {\bibfnamefont {M.}~\bibnamefont {Sato}}\ and\ \bibinfo {author} {\bibfnamefont {S.}~\bibnamefont {Fujimoto}},\ }\bibfield  {title} {\bibinfo {title} {Majorana fermions and topology in superconductors},\ }\href@noop {} {\bibfield  {journal} {\bibinfo  {journal} {Journal of the Physical Society of Japan}\ }\textbf {\bibinfo {volume} {85}},\ \bibinfo {pages} {072001} (\bibinfo {year} {2016})}\BibitemShut {NoStop}%
\bibitem [{\citenamefont {Qi}\ and\ \citenamefont {Zhang}(2011)}]{qi2011topological}%
  \BibitemOpen
  \bibfield  {author} {\bibinfo {author} {\bibfnamefont {X.-L.}\ \bibnamefont {Qi}}\ and\ \bibinfo {author} {\bibfnamefont {S.-C.}\ \bibnamefont {Zhang}},\ }\bibfield  {title} {\bibinfo {title} {Topological insulators and superconductors},\ }\href@noop {} {\bibfield  {journal} {\bibinfo  {journal} {Reviews of modern physics}\ }\textbf {\bibinfo {volume} {83}},\ \bibinfo {pages} {1057} (\bibinfo {year} {2011})}\BibitemShut {NoStop}%
\bibitem [{\citenamefont {Tsutsumi}\ \emph {et~al.}(2008)\citenamefont {Tsutsumi}, \citenamefont {Kawakami}, \citenamefont {Mizushima}, \citenamefont {Ichioka},\ and\ \citenamefont {Machida}}]{tsutsumi2008majorana}%
  \BibitemOpen
  \bibfield  {author} {\bibinfo {author} {\bibfnamefont {Y.}~\bibnamefont {Tsutsumi}}, \bibinfo {author} {\bibfnamefont {T.}~\bibnamefont {Kawakami}}, \bibinfo {author} {\bibfnamefont {T.}~\bibnamefont {Mizushima}}, \bibinfo {author} {\bibfnamefont {M.}~\bibnamefont {Ichioka}},\ and\ \bibinfo {author} {\bibfnamefont {K.}~\bibnamefont {Machida}},\ }\bibfield  {title} {\bibinfo {title} {Majorana bound state in rotating superfluid he 3- a between parallel plates},\ }\href@noop {} {\bibfield  {journal} {\bibinfo  {journal} {Physical review letters}\ }\textbf {\bibinfo {volume} {101}},\ \bibinfo {pages} {135302} (\bibinfo {year} {2008})}\BibitemShut {NoStop}%
\bibitem [{\citenamefont {Moroz}\ and\ \citenamefont {Hoyos}(2015)}]{moroz2015effective}%
  \BibitemOpen
  \bibfield  {author} {\bibinfo {author} {\bibfnamefont {S.}~\bibnamefont {Moroz}}\ and\ \bibinfo {author} {\bibfnamefont {C.}~\bibnamefont {Hoyos}},\ }\bibfield  {title} {\bibinfo {title} {Effective theory of two-dimensional chiral superfluids: gauge duality and newton-cartan formulation},\ }\href@noop {} {\bibfield  {journal} {\bibinfo  {journal} {Physical Review B}\ }\textbf {\bibinfo {volume} {91}},\ \bibinfo {pages} {064508} (\bibinfo {year} {2015})}\BibitemShut {NoStop}%
\bibitem [{\citenamefont {Kvorning}\ \emph {et~al.}(2018)\citenamefont {Kvorning}, \citenamefont {Hansson}, \citenamefont {Quelle},\ and\ \citenamefont {Smith}}]{kvorning2018proposed}%
  \BibitemOpen
  \bibfield  {author} {\bibinfo {author} {\bibfnamefont {T.}~\bibnamefont {Kvorning}}, \bibinfo {author} {\bibfnamefont {T.~H.}\ \bibnamefont {Hansson}}, \bibinfo {author} {\bibfnamefont {A.}~\bibnamefont {Quelle}},\ and\ \bibinfo {author} {\bibfnamefont {C.~M.}\ \bibnamefont {Smith}},\ }\bibfield  {title} {\bibinfo {title} {Proposed spontaneous generation of magnetic fields by curved layers of a chiral superconductor},\ }\href@noop {} {\bibfield  {journal} {\bibinfo  {journal} {Physical Review Letters}\ }\textbf {\bibinfo {volume} {120}},\ \bibinfo {pages} {217002} (\bibinfo {year} {2018})}\BibitemShut {NoStop}%
\bibitem [{\citenamefont {Jiang}\ \emph {et~al.}(2020)\citenamefont {Jiang}, \citenamefont {Hansson},\ and\ \citenamefont {Wilczek}}]{jiang2020geometric}%
  \BibitemOpen
  \bibfield  {author} {\bibinfo {author} {\bibfnamefont {Q.-D.}\ \bibnamefont {Jiang}}, \bibinfo {author} {\bibfnamefont {T.~H.}\ \bibnamefont {Hansson}},\ and\ \bibinfo {author} {\bibfnamefont {F.}~\bibnamefont {Wilczek}},\ }\bibfield  {title} {\bibinfo {title} {Geometric induction in chiral superconductors},\ }\href@noop {} {\bibfield  {journal} {\bibinfo  {journal} {Physical Review Letters}\ }\textbf {\bibinfo {volume} {124}},\ \bibinfo {pages} {197001} (\bibinfo {year} {2020})}\BibitemShut {NoStop}%
\bibitem [{\citenamefont {Jiang}\ and\ \citenamefont {Balatsky}(2022)}]{qingdong}%
  \BibitemOpen
  \bibfield  {author} {\bibinfo {author} {\bibfnamefont {Q.-D.}\ \bibnamefont {Jiang}}\ and\ \bibinfo {author} {\bibfnamefont {A.}~\bibnamefont {Balatsky}},\ }\bibfield  {title} {\bibinfo {title} {Geometric induction in chiral superfluids},\ }\href {https://doi.org/10.1103/PhysRevLett.129.016801} {\bibfield  {journal} {\bibinfo  {journal} {Phys. Rev. Lett.}\ }\textbf {\bibinfo {volume} {129}},\ \bibinfo {pages} {016801} (\bibinfo {year} {2022})}\BibitemShut {NoStop}%
\bibitem [{\citenamefont {Bowick}\ and\ \citenamefont {Travesset}(2001)}]{bowick2001statistical}%
  \BibitemOpen
  \bibfield  {author} {\bibinfo {author} {\bibfnamefont {M.~J.}\ \bibnamefont {Bowick}}\ and\ \bibinfo {author} {\bibfnamefont {A.}~\bibnamefont {Travesset}},\ }\bibfield  {title} {\bibinfo {title} {The statistical mechanics of membranes},\ }\href@noop {} {\bibfield  {journal} {\bibinfo  {journal} {Physics Reports}\ }\textbf {\bibinfo {volume} {344}},\ \bibinfo {pages} {255} (\bibinfo {year} {2001})}\BibitemShut {NoStop}%
\bibitem [{\citenamefont {Nelson}\ \emph {et~al.}(2004)\citenamefont {Nelson}, \citenamefont {Piran},\ and\ \citenamefont {Weinberg}}]{nelson2004statistical}%
  \BibitemOpen
  \bibfield  {author} {\bibinfo {author} {\bibfnamefont {D.}~\bibnamefont {Nelson}}, \bibinfo {author} {\bibfnamefont {T.}~\bibnamefont {Piran}},\ and\ \bibinfo {author} {\bibfnamefont {S.}~\bibnamefont {Weinberg}},\ }\href@noop {} {\emph {\bibinfo {title} {Statistical mechanics of membranes and surfaces}}}\ (\bibinfo  {publisher} {World Scientific},\ \bibinfo {year} {2004})\BibitemShut {NoStop}%
\bibitem [{\citenamefont {Helfrich}(1973)}]{helfrich1973elastic}%
  \BibitemOpen
  \bibfield  {author} {\bibinfo {author} {\bibfnamefont {W.}~\bibnamefont {Helfrich}},\ }\bibfield  {title} {\bibinfo {title} {Elastic properties of lipid bilayers: theory and possible experiments},\ }\href@noop {} {\bibfield  {journal} {\bibinfo  {journal} {Zeitschrift f{\"u}r Naturforschung c}\ }\textbf {\bibinfo {volume} {28}},\ \bibinfo {pages} {693} (\bibinfo {year} {1973})}\BibitemShut {NoStop}%
\bibitem [{\citenamefont {De~Gennes}\ and\ \citenamefont {Taupin}(1982)}]{1982degennes}%
  \BibitemOpen
  \bibfield  {author} {\bibinfo {author} {\bibfnamefont {P.}~\bibnamefont {De~Gennes}}\ and\ \bibinfo {author} {\bibfnamefont {C.}~\bibnamefont {Taupin}},\ }\bibfield  {title} {\bibinfo {title} {Microemulsions and the flexibility of oil/water interfaces},\ }\href@noop {} {\bibfield  {journal} {\bibinfo  {journal} {The Journal of physical chemistry}\ }\textbf {\bibinfo {volume} {86}},\ \bibinfo {pages} {2294} (\bibinfo {year} {1982})}\BibitemShut {NoStop}%
\bibitem [{\citenamefont {Nelson}\ and\ \citenamefont {Peliti}(1987)}]{nelson1987fluctuations}%
  \BibitemOpen
  \bibfield  {author} {\bibinfo {author} {\bibfnamefont {D.}~\bibnamefont {Nelson}}\ and\ \bibinfo {author} {\bibfnamefont {L.}~\bibnamefont {Peliti}},\ }\bibfield  {title} {\bibinfo {title} {Fluctuations in membranes with crystalline and hexatic order},\ }\href@noop {} {\bibfield  {journal} {\bibinfo  {journal} {Journal de physique}\ }\textbf {\bibinfo {volume} {48}},\ \bibinfo {pages} {1085} (\bibinfo {year} {1987})}\BibitemShut {NoStop}%
\bibitem [{\citenamefont {Guitter}\ and\ \citenamefont {Kardar}(1990)}]{guitter1990tethering}%
  \BibitemOpen
  \bibfield  {author} {\bibinfo {author} {\bibfnamefont {E.}~\bibnamefont {Guitter}}\ and\ \bibinfo {author} {\bibfnamefont {M.}~\bibnamefont {Kardar}},\ }\bibfield  {title} {\bibinfo {title} {Tethering, crumpling, and melting transitions in hexatic membranes},\ }\href@noop {} {\bibfield  {journal} {\bibinfo  {journal} {Europhysics Letters}\ }\textbf {\bibinfo {volume} {13}},\ \bibinfo {pages} {441} (\bibinfo {year} {1990})}\BibitemShut {NoStop}%
\bibitem [{\citenamefont {Shankar}\ and\ \citenamefont {Nelson}(2021)}]{shankar2021thermalized}%
  \BibitemOpen
  \bibfield  {author} {\bibinfo {author} {\bibfnamefont {S.}~\bibnamefont {Shankar}}\ and\ \bibinfo {author} {\bibfnamefont {D.~R.}\ \bibnamefont {Nelson}},\ }\bibfield  {title} {\bibinfo {title} {Thermalized buckling of isotropically compressed thin sheets},\ }\href@noop {} {\bibfield  {journal} {\bibinfo  {journal} {Physical Review E}\ }\textbf {\bibinfo {volume} {104}},\ \bibinfo {pages} {054141} (\bibinfo {year} {2021})}\BibitemShut {NoStop}%
\bibitem [{\citenamefont {Metayer}\ \emph {et~al.}(2022)\citenamefont {Metayer}, \citenamefont {Mouhanna},\ and\ \citenamefont {Teber}}]{metayer2022three}%
  \BibitemOpen
  \bibfield  {author} {\bibinfo {author} {\bibfnamefont {S.}~\bibnamefont {Metayer}}, \bibinfo {author} {\bibfnamefont {D.}~\bibnamefont {Mouhanna}},\ and\ \bibinfo {author} {\bibfnamefont {S.}~\bibnamefont {Teber}},\ }\bibfield  {title} {\bibinfo {title} {Three-loop order approach to flat polymerized membranes},\ }\href@noop {} {\bibfield  {journal} {\bibinfo  {journal} {Physical Review E}\ }\textbf {\bibinfo {volume} {105}},\ \bibinfo {pages} {L012603} (\bibinfo {year} {2022})}\BibitemShut {NoStop}%
\bibitem [{\citenamefont {Nicholl}\ \emph {et~al.}(2017)\citenamefont {Nicholl}, \citenamefont {Lavrik}, \citenamefont {Vlassiouk}, \citenamefont {Srijanto},\ and\ \citenamefont {Bolotin}}]{nicholl2017hidden}%
  \BibitemOpen
  \bibfield  {author} {\bibinfo {author} {\bibfnamefont {R.~J.}\ \bibnamefont {Nicholl}}, \bibinfo {author} {\bibfnamefont {N.~V.}\ \bibnamefont {Lavrik}}, \bibinfo {author} {\bibfnamefont {I.}~\bibnamefont {Vlassiouk}}, \bibinfo {author} {\bibfnamefont {B.~R.}\ \bibnamefont {Srijanto}},\ and\ \bibinfo {author} {\bibfnamefont {K.~I.}\ \bibnamefont {Bolotin}},\ }\bibfield  {title} {\bibinfo {title} {Hidden area and mechanical nonlinearities in freestanding graphene},\ }\href@noop {} {\bibfield  {journal} {\bibinfo  {journal} {Physical review letters}\ }\textbf {\bibinfo {volume} {118}},\ \bibinfo {pages} {266101} (\bibinfo {year} {2017})}\BibitemShut {NoStop}%
\bibitem [{\citenamefont {Lopez-Polin}\ \emph {et~al.}(2022)\citenamefont {Lopez-Polin}, \citenamefont {Gomez-Navarro},\ and\ \citenamefont {Gomez-Herrero}}]{lopez2022effect}%
  \BibitemOpen
  \bibfield  {author} {\bibinfo {author} {\bibfnamefont {G.}~\bibnamefont {Lopez-Polin}}, \bibinfo {author} {\bibfnamefont {C.}~\bibnamefont {Gomez-Navarro}},\ and\ \bibinfo {author} {\bibfnamefont {J.}~\bibnamefont {Gomez-Herrero}},\ }\bibfield  {title} {\bibinfo {title} {The effect of rippling on the mechanical properties of graphene},\ }\href@noop {} {\bibfield  {journal} {\bibinfo  {journal} {Nano Materials Science}\ }\textbf {\bibinfo {volume} {4}},\ \bibinfo {pages} {18} (\bibinfo {year} {2022})}\BibitemShut {NoStop}%
\bibitem [{sup()}]{supplementary}%
  \BibitemOpen
  \href@noop {} {\bibinfo {title} {See supplementary material}}\BibitemShut {NoStop}%
\bibitem [{\citenamefont {Chui}\ and\ \citenamefont {Lee}(1975)}]{chuilee1975}%
  \BibitemOpen
  \bibfield  {author} {\bibinfo {author} {\bibfnamefont {S.~T.}\ \bibnamefont {Chui}}\ and\ \bibinfo {author} {\bibfnamefont {P.~A.}\ \bibnamefont {Lee}},\ }\bibfield  {title} {\bibinfo {title} {Equivalence of a one-dimensional fermion model and the two-dimensional coulomb gas},\ }\href {https://doi.org/10.1103/PhysRevLett.35.315} {\bibfield  {journal} {\bibinfo  {journal} {Phys. Rev. Lett.}\ }\textbf {\bibinfo {volume} {35}},\ \bibinfo {pages} {315} (\bibinfo {year} {1975})}\BibitemShut {NoStop}%
\bibitem [{\citenamefont {Turner}\ \emph {et~al.}(2010)\citenamefont {Turner}, \citenamefont {Vitelli},\ and\ \citenamefont {Nelson}}]{turner2010vortices}%
  \BibitemOpen
  \bibfield  {author} {\bibinfo {author} {\bibfnamefont {A.~M.}\ \bibnamefont {Turner}}, \bibinfo {author} {\bibfnamefont {V.}~\bibnamefont {Vitelli}},\ and\ \bibinfo {author} {\bibfnamefont {D.~R.}\ \bibnamefont {Nelson}},\ }\bibfield  {title} {\bibinfo {title} {Vortices on curved surfaces},\ }\href@noop {} {\bibfield  {journal} {\bibinfo  {journal} {Reviews of Modern Physics}\ }\textbf {\bibinfo {volume} {82}},\ \bibinfo {pages} {1301} (\bibinfo {year} {2010})}\BibitemShut {NoStop}%
\bibitem [{\citenamefont {Wiegmann}\ and\ \citenamefont {Abanov}(2014)}]{wiegmann2014anomalous}%
  \BibitemOpen
  \bibfield  {author} {\bibinfo {author} {\bibfnamefont {P.}~\bibnamefont {Wiegmann}}\ and\ \bibinfo {author} {\bibfnamefont {A.~G.}\ \bibnamefont {Abanov}},\ }\bibfield  {title} {\bibinfo {title} {Anomalous hydrodynamics of two-dimensional vortex fluids},\ }\href@noop {} {\bibfield  {journal} {\bibinfo  {journal} {Physical review letters}\ }\textbf {\bibinfo {volume} {113}},\ \bibinfo {pages} {034501} (\bibinfo {year} {2014})}\BibitemShut {NoStop}%
\bibitem [{\citenamefont {Kosterlitz}\ and\ \citenamefont {Thouless}(2002)}]{KT}%
  \BibitemOpen
  \bibfield  {author} {\bibinfo {author} {\bibfnamefont {J.~M.}\ \bibnamefont {Kosterlitz}}\ and\ \bibinfo {author} {\bibfnamefont {D.~J.}\ \bibnamefont {Thouless}},\ }\bibfield  {title} {\bibinfo {title} {{Ordering, metastability and phase transitions in two-dimensional systems}},\ }\href {https://doi.org/10.1088/0022-3719/6/7/010} {\bibfield  {journal} {\bibinfo  {journal} {Journal of Physics C: Solid State Physics}\ }\textbf {\bibinfo {volume} {6}},\ \bibinfo {pages} {1181} (\bibinfo {year} {2002})}\BibitemShut {NoStop}%
\bibitem [{\citenamefont {Shin}(1991)}]{riemannsurface}%
  \BibitemOpen
  \bibfield  {author} {\bibinfo {author} {\bibfnamefont {H.~J.}\ \bibnamefont {Shin}},\ }\bibfield  {title} {\bibinfo {title} {Coulomb-gas representation of minimal models on riemann surfaces},\ }\href {https://doi.org/10.1103/PhysRevD.44.3843} {\bibfield  {journal} {\bibinfo  {journal} {Phys. Rev. D}\ }\textbf {\bibinfo {volume} {44}},\ \bibinfo {pages} {3843} (\bibinfo {year} {1991})}\BibitemShut {NoStop}%
\bibitem [{\citenamefont {Coleman}(1975)}]{coleman75sinegordon}%
  \BibitemOpen
  \bibfield  {author} {\bibinfo {author} {\bibfnamefont {S.}~\bibnamefont {Coleman}},\ }\bibfield  {title} {\bibinfo {title} {Quantum sine-gordon equation as the massive thirring model},\ }\href {https://doi.org/10.1103/PhysRevD.11.2088} {\bibfield  {journal} {\bibinfo  {journal} {Phys. Rev. D}\ }\textbf {\bibinfo {volume} {11}},\ \bibinfo {pages} {2088} (\bibinfo {year} {1975})}\BibitemShut {NoStop}%
\bibitem [{\citenamefont {David}\ \emph {et~al.}(1987)\citenamefont {David}, \citenamefont {Guitter},\ and\ \citenamefont {Peliti}}]{david1987critical}%
  \BibitemOpen
  \bibfield  {author} {\bibinfo {author} {\bibfnamefont {F.}~\bibnamefont {David}}, \bibinfo {author} {\bibfnamefont {E.}~\bibnamefont {Guitter}},\ and\ \bibinfo {author} {\bibfnamefont {L.}~\bibnamefont {Peliti}},\ }\bibfield  {title} {\bibinfo {title} {Critical properties of fluid membranes with hexatic order},\ }\href@noop {} {\bibfield  {journal} {\bibinfo  {journal} {Journal de Physique}\ }\textbf {\bibinfo {volume} {48}},\ \bibinfo {pages} {2059} (\bibinfo {year} {1987})}\BibitemShut {NoStop}%
\bibitem [{\citenamefont {Park}\ and\ \citenamefont {Lubensky}(1996{\natexlab{a}})}]{KTpaticsurface}%
  \BibitemOpen
  \bibfield  {author} {\bibinfo {author} {\bibfnamefont {J.-M.}\ \bibnamefont {Park}}\ and\ \bibinfo {author} {\bibfnamefont {T.~C.}\ \bibnamefont {Lubensky}},\ }\bibfield  {title} {\bibinfo {title} {Topological defects on fluctuating surfaces: General properties and the kosterlitz-thouless transition},\ }\href {https://doi.org/10.1103/PhysRevE.53.2648} {\bibfield  {journal} {\bibinfo  {journal} {Phys. Rev. E}\ }\textbf {\bibinfo {volume} {53}},\ \bibinfo {pages} {2648} (\bibinfo {year} {1996}{\natexlab{a}})}\BibitemShut {NoStop}%
\bibitem [{\citenamefont {Park}\ and\ \citenamefont {Lubensky}(1996{\natexlab{b}})}]{park1996sine}%
  \BibitemOpen
  \bibfield  {author} {\bibinfo {author} {\bibfnamefont {J.-M.}\ \bibnamefont {Park}}\ and\ \bibinfo {author} {\bibfnamefont {T.}~\bibnamefont {Lubensky}},\ }\bibfield  {title} {\bibinfo {title} {Sine-gordon field theory for the kosterlitz-thouless transitions on fluctuating membranes},\ }\href@noop {} {\bibfield  {journal} {\bibinfo  {journal} {Physical Review E}\ }\textbf {\bibinfo {volume} {53}},\ \bibinfo {pages} {2665} (\bibinfo {year} {1996}{\natexlab{b}})}\BibitemShut {NoStop}%
\bibitem [{\citenamefont {David}\ and\ \citenamefont {Guitter}(1988)}]{david1988crumpling}%
  \BibitemOpen
  \bibfield  {author} {\bibinfo {author} {\bibfnamefont {F.}~\bibnamefont {David}}\ and\ \bibinfo {author} {\bibfnamefont {E.}~\bibnamefont {Guitter}},\ }\bibfield  {title} {\bibinfo {title} {Crumpling transition in elastic membranes: renormalization group treatment},\ }\href@noop {} {\bibfield  {journal} {\bibinfo  {journal} {Europhysics Letters}\ }\textbf {\bibinfo {volume} {5}},\ \bibinfo {pages} {709} (\bibinfo {year} {1988})}\BibitemShut {NoStop}%
\bibitem [{\citenamefont {Aronovitz}\ and\ \citenamefont {Lubensky}(1988)}]{aronovitz1988fluctuations}%
  \BibitemOpen
  \bibfield  {author} {\bibinfo {author} {\bibfnamefont {J.~A.}\ \bibnamefont {Aronovitz}}\ and\ \bibinfo {author} {\bibfnamefont {T.~C.}\ \bibnamefont {Lubensky}},\ }\bibfield  {title} {\bibinfo {title} {Fluctuations of solid membranes},\ }\href@noop {} {\bibfield  {journal} {\bibinfo  {journal} {Physical review letters}\ }\textbf {\bibinfo {volume} {60}},\ \bibinfo {pages} {2634} (\bibinfo {year} {1988})}\BibitemShut {NoStop}%
\bibitem [{\citenamefont {Le~Doussal}\ and\ \citenamefont {Radzihovsky}(1992)}]{le1992self}%
  \BibitemOpen
  \bibfield  {author} {\bibinfo {author} {\bibfnamefont {P.}~\bibnamefont {Le~Doussal}}\ and\ \bibinfo {author} {\bibfnamefont {L.}~\bibnamefont {Radzihovsky}},\ }\bibfield  {title} {\bibinfo {title} {Self-consistent theory of polymerized membranes},\ }\href@noop {} {\bibfield  {journal} {\bibinfo  {journal} {Physical review letters}\ }\textbf {\bibinfo {volume} {69}},\ \bibinfo {pages} {1209} (\bibinfo {year} {1992})}\BibitemShut {NoStop}%
\bibitem [{\citenamefont {Mauri}\ and\ \citenamefont {Katsnelson}(2020)}]{mauri2020scaling}%
  \BibitemOpen
  \bibfield  {author} {\bibinfo {author} {\bibfnamefont {A.}~\bibnamefont {Mauri}}\ and\ \bibinfo {author} {\bibfnamefont {M.~I.}\ \bibnamefont {Katsnelson}},\ }\bibfield  {title} {\bibinfo {title} {Scaling behavior of crystalline membranes: An $\varepsilon$-expansion approach},\ }\href@noop {} {\bibfield  {journal} {\bibinfo  {journal} {Nuclear Physics B}\ }\textbf {\bibinfo {volume} {956}},\ \bibinfo {pages} {115040} (\bibinfo {year} {2020})}\BibitemShut {NoStop}%
\bibitem [{\citenamefont {Amit}\ \emph {et~al.}(1980)\citenamefont {Amit}, \citenamefont {Goldschmidt},\ and\ \citenamefont {Grinstein}}]{amit1980renormalisation}%
  \BibitemOpen
  \bibfield  {author} {\bibinfo {author} {\bibfnamefont {D.~J.}\ \bibnamefont {Amit}}, \bibinfo {author} {\bibfnamefont {Y.~Y.}\ \bibnamefont {Goldschmidt}},\ and\ \bibinfo {author} {\bibfnamefont {S.}~\bibnamefont {Grinstein}},\ }\bibfield  {title} {\bibinfo {title} {Renormalisation group analysis of the phase transition in the 2d coulomb gas, sine-gordon theory and xy-model},\ }\href@noop {} {\bibfield  {journal} {\bibinfo  {journal} {Journal of Physics A: Mathematical and General}\ }\textbf {\bibinfo {volume} {13}},\ \bibinfo {pages} {585} (\bibinfo {year} {1980})}\BibitemShut {NoStop}%
\bibitem [{\citenamefont {Wiegmann}(1978)}]{wiegmann1978one}%
  \BibitemOpen
  \bibfield  {author} {\bibinfo {author} {\bibfnamefont {P.}~\bibnamefont {Wiegmann}},\ }\bibfield  {title} {\bibinfo {title} {One-dimensional fermi system and plane xy model},\ }\href@noop {} {\bibfield  {journal} {\bibinfo  {journal} {Journal of Physics C: Solid State Physics}\ }\textbf {\bibinfo {volume} {11}},\ \bibinfo {pages} {1583} (\bibinfo {year} {1978})}\BibitemShut {NoStop}%
\end{thebibliography}%

\widetext
\newpage
\begin{center}
\textbf{\large Supplemental Material: Geometry fluctuations in chiral superfluids}
\end{center}
\setcounter{equation}{0}
\setcounter{figure}{0}
\setcounter{table}{0}
\makeatletter
\renewcommand{\theequation}{S\arabic{equation}}
\renewcommand{\thefigure}{S\arabic{figure}}
\renewcommand{\bibnumfmt}[1]{[S#1]}
\renewcommand{\citenumfont}[1]{S#1}

\section{I. 2+1-dimensional geometry}

In a local basis for the tangent space $e_{1,2}$, one can express the order parameter of a chiral superfluid with angular momentum $\ell$ as
\begin{equation}
    \Psi=\psi \epsilon^\ell_{\pm},
\end{equation}
where
\begin{equation}
    \epsilon^\ell_{\pm}= \otimes_{j=1}^\ell\frac{(e_1\pm e_2)}{2},
\end{equation}
gives a basis for positive/negative chirality configurations and $\psi=\sqrt{\rho}e^{i\theta}$ is a locally defined wavefunction. Then the covariant derivative becomes
\begin{equation}
    D_a(\psi \epsilon_\pm^\ell)=(\p_a\pm \ell\omega_a)\psi\epsilon_\pm^\ell,
\end{equation}
where $\omega_a=e_1^\dagger\p_a e_2$ is the spin connection. In principle, the local basis vectors are time-dependent, leading to a temporal component of the spin connection. In our notation, latin indices of the begining of the alphabet $a,b,...$ represent spacetime indices with Euclidean signature, while $i,j,...$ are spatial indices.

The geometric configurations we consider are the different embeddings $\mathbf{r}\mapsto\mathbf{R}$ of the membrane into three-dimensional space. The induced metrics are of the form
\begin{equation}
    \begin{pmatrix}
        1&\\
        &g_{ij}
    \end{pmatrix},\label{eq:gmatrix}
\end{equation}
where $g_{ij}=\p_i\vec{r}\cdot\p_j\vec{r}_j$ is static,
\begin{equation}
    \partial_t g_{ij}=0.\label{eq:dtg}
\end{equation}

The coupling to the superfluid degrees of freedom is through the geometric current,
\begin{equation}
    j_g^a=\ell\epsilon^{abc}\p_b\omega_c.
\end{equation}
Since the spin-connection simplifies to $\vec{\omega}=(\omega_1,\omega_2)(\mathbf{r})$, it follows that the spatial part vanishes,
\begin{equation}
    \vec{j}_g=(\partial_2\omega_0-\partial_0\omega_2,\partial_0\omega_1-\partial_1\omega_0)=0,\label{eq:jspatial}
\end{equation}
and the coupling of the membrane geometry to the superfluid is thus contained in the time-component, which is given by the Gaussian curvature,
\begin{equation}
    j_g^0=\epsilon^{ij}\p_i\omega_j= R.\label{eq:j0R}
\end{equation}
This means that, although the bending energy depends on the embedding geometry (ie. the mean curvature), the superfluid is only sensitive to the intrinsic geometry of the membrane (ie. the Gaussian curvature).

\subsection{The Monge parametrization}

On a nearly flat membrane, the embedding map can be expressed in the Monge parametrization by $\vec{r}(x,y)=(x,y,h(x,y))$, with $h$ small and smooth. Then one can easily verify that, to lowest order in gradients,
\begin{equation}
    (g_{ij})=(\p_i \vec{r}\cdot\p_j \vec{r})=
    \begin{pmatrix}
        1+(\p_1 h)^2 & \p_1 h\p_2 h\\
        \p_1 h\p_2 h & 1+(\p_2 h)^2
    \end{pmatrix},
    g=1+(\p h)^2, (g^{ij})=\frac{1}{g}
    \begin{pmatrix}
        g_{22} & -g_{21}\\
        -g_{12} & g_{11}
    \end{pmatrix},\label{eq:gformulas1}
\end{equation}
\begin{align}
    &\vec{n}=\frac{\p_1\vec{r}\times \p_2\vec{r}}{|\p_1\vec{r}\times \p_2\vec{r}|}=\frac{(-\p_1h,-\p_2h,1)}{\sqrt{1+(\nabla h)^2}}, &(K_{ij}):=(\vec{n}\cdot \p_i \p_j \vec{r})=\frac{1}{\sqrt{1+(\nabla h)^2}}
    \begin{pmatrix}
        \p^2_1h & \p^2_{12}h\\
        \p^2_{12}h & \p^2_2h
    \end{pmatrix},\label{eq:gformulas2}
\end{align}
\begin{align}
    H:=&g^{ij}K_{ij}=\Delta h+O(h^3),\\
    R:=&\det(g^{ij}K_{jk})=\frac{1}{g}\det K=\p^2_1h\p^2_2h-\p^2_{12}h\p^2_{12}h+O(h^4)=\frac{1}{2}[(\Delta h)^2-(\p^2_{ij}h)^2]+O(h^4).\label{eq:Rh}
\end{align}
When considering the renormalization of the geometry propagator, the next order term in the expansion of the mean curvature,
\begin{align}
    \sqrt{g}H^2=&\sqrt{g}(g^{\mu\nu}K_{\mu\nu})^2=\frac{1}{g^{\frac{5}{2}}}(\Delta h+\bp^\mu h\p^2_{\mu\nu}h\bp^\nu h)^2=[1+(\nabla h)^2]^{-\frac{5}{2}}(\Delta h)^2\left[1+\frac{\bp^\mu h\p^2_{\mu\nu}h\bp^\nu h}{\Delta h}\right]^2\nonumber\\
    =&(\Delta h)^2-\frac{5}{2}(\Delta h)^2(\nabla h)^2+2\Delta h \bp^\mu h\p^2_{\mu\nu}h\bp^\nu h+O(h^6),
\end{align}
leads to the vertex
\begin{align}
    \frac{2}{\tilde{\kappa}}V_h = -\frac{5}{2}(\Delta h)^2(\nabla h)^2+2\Delta h \bp^i h\p^2_{ij}h\bp^j h,
\end{align}
which contributes to the renormalization of the bending rigidity (see below). We use the convenient notation $\bar{\p}_i=\epsilon_{ij}\p_j$

\subsection{The Riemann tensor}

The metrics (\ref{eq:gmatrix},\ref{eq:gformulas1}) can generate Riemannian curvature. We find that the Riemann tensor has only four non-zero components,
\begin{align}
    R_{2323}=-R_{2332}=-R_{3223}=R_{3232}=R,\label{eq:Riemanntensor}
\end{align}
where $R$ is the Gaussian curvature (\ref{eq:Rh}). A consequence of this highly symmetric solution appears when extending our process of integrating out the smooth part of the phase field $\theta_s$ to the dynamical case. Then the action has a term
\begin{equation}
    \int d^2xdt\xi^a\p_a\theta_s,
\end{equation}
so that integrating out $\theta_s$ leads to the equation
\begin{equation}
    \p_a(\sqg\xi^a)=0\Rightarrow D_a\xi^a=0.
\end{equation}
In flat space, one solves this equation by introducing a potential field
\begin{equation}
    \xi^a=\epsilon^{abc}\p_b A_c,
\end{equation}
and using the commutation of partial derivatives. If $\xi^a$ is smooth and single-valued, then all solutions can be expressed in this form, by Poincar\'e's lemma. In curved geometry, one can consider
\begin{equation}
    \xi^a=\epsilon^{abc}D_b A_c.\label{eq:curveDA}
\end{equation}
This is in general not a solution because covariant derivatives do not commute. In fact, the definition of the Riemann curvature tensor is
\begin{equation}
    (D_aD_b-D_bD_a)A_c=R^d_{abc} A_d.
\end{equation}
Instead, we make attempt at an expansion in curvature, by considering the shift
\begin{equation}
    \xi^a=\epsilon^{abc}D_b A_c+v^a,
\end{equation}
where $A_c$ is a general vector field. Then $v^a$ should be a solution of the equation
\begin{equation}
    D_av^a=-\frac{1}{2}\epsilon^{abc}R^{d}_{abc}A_d.
\end{equation}
The right-hand side vanishes for the $2+1$-dimensional membrane configurations, since all coefficients of $R^d_{abc}$ involving time components are zero (\ref{eq:Riemanntensor}). Therefore, we can take $v^a=0$.

\newpage

\section{II. Background tension}

As discussed in the main text, the presence of tension modifies the amplitude of geometry fluctuations at large distances. This is most clear from the height field propagator,
\begin{equation}
    G_h(q)=\frac{1}{\tilde{\sigma} q^2+\tilde{\kappa} q^4}.
\end{equation}
In terms of the momentum scale
\begin{equation}
    q_\sigma = \sqrt{\frac{\sigma}{\kappa}},
\end{equation}
the propagator is dominated by the tension term for $q<q_\sigma$ and by the bending term for $q>q_\sigma$. Thus one can approximately evaluate the loop integral as
\begin{align}
    \langle R_h(q) R_h(-q)\rangle_h &=\int\frac{d^2l}{(2\pi)^2}\frac{(q\times l)^4}{8\pi^2}G_h(l)G_h(q-l) \\
    &\approx \int_{\frac{\pi}{L}<|l|<q_\sigma}\frac{d^2l}{(2\pi)^2}\frac{(q\times l)^4}{8\pi^2}\frac{1}{\tilde{\sigma}l^2\tilde{\sigma}(q-l)^2} + \int_{q_\sigma<|l|<\frac{\pi}{a}}\frac{d^2l}{(2\pi)^2}\frac{(q\times l)^4}{8\pi^2}\frac{1}{\tilde{\kappa}l^4\tilde{\kappa}(q-l)^4}.\label{eq:cutoffsigmakappa}
\end{align}
Note that in the small tension limit, $q_\sigma\to 0$, one has
\begin{align}
    \langle R_h(q) R_h(-q)\rangle_h &= \int_{\frac{\pi}{L}<|l|<\frac{\pi}{a}}\frac{d^2l}{(2\pi)^2}\frac{(q\times l)^4}{8\pi^2}\frac{1}{\tilde{\kappa}l^4\tilde{\kappa}(q-l)^4} \\
    &=\frac{3q^2}{16(2\pi)^3}\tilde{\kappa}^{-2},
\end{align}
where in the last step we kept the leading term in $a$, $1/L$. As we discussed in the main text, this result modifies the vortex-antivortex confining potential at long distances, leading to a change in the energy scale prefactor. On the other hand, in the large tension limit $q_\sigma\to\infty$, one has
\begin{align}
    \langle R_h(q) R_h(-q)\rangle_h &= \int_{\frac{\pi}{L}<|l|<\frac{\pi}{a}}\frac{d^2l}{(2\pi)^2}\frac{(q\times l)^4}{8\pi^2}\frac{1}{\tilde{\sigma}l^2\tilde{\sigma}(q-l)^2}\\
    &=\frac{3q^4}{32(2\pi)^3}\left(\frac{\pi}{a}\right)^2\tilde{\sigma}^{-2},
\end{align}
where we note the ultraviolet divergence. Importantly, this evaluates to a $q^4$ contribution. The effective potential becomes
\begin{equation}
    \Gamma[\phi] = \int d^2x \left[\frac{(\p\phi)^2}{2\tilde{\gamma}}+\frac{4\pi}{32}\left(\frac{\pi}{a}\right)^2\tilde{\sigma}^{-2}(\Delta\phi)^2+i\phi\rho_v\right],
\end{equation}
and the resulting energy of a vortex pair,
\begin{equation}
    E = 2E_C+2\pi\gamma\left(\log\Big|\frac{r_1-r_2}{a}\Big|+K_0\Big|\frac{r_1-r_2}{\epsilon_\sigma}\Big|\right),
\end{equation}
with
\begin{equation}
    \epsilon_\sigma^2=\frac{3\pi}{16}\left(\frac{\pi}{a}\right)^2\frac{\gamma T}{\sigma^2}.
\end{equation}
The modified Bessel function of the second kind $K_0(x)$ vanishes exponentially fast at large $x$, so that the vortex-antivortex confining potential has the same logarithmic growth at large distances as in flat space, and in fact with the same prefactor $\gamma$. The effect of the large tension is to suppress this interaction at scales smaller than $\epsilon_\sigma\propto\sigma^{-1}$, cancelling out the UV divergence of the logarithmic potential.

\begin{figure}[h!]
    \centering 
    \includegraphics[width=0.55\textwidth]{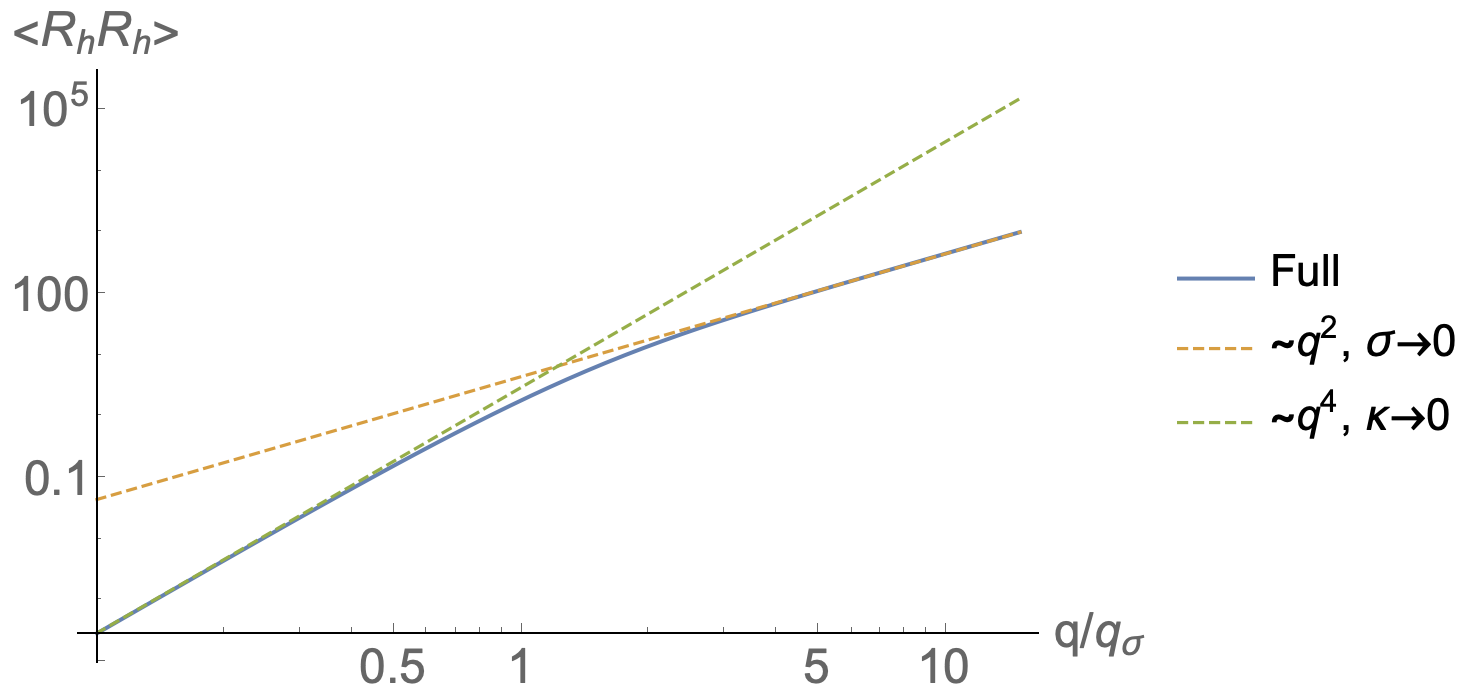}
    \caption{Log-log plot of the two-point correlation function of the curvature as a function of the normalized momentum $q/q_\sigma$. Around $q\sim q_\sigma$, the dependence on the momentum changes for quadratic to quartic. This corresponds to the transition from the large tension to the small tension regime.}
    \label{fig:RRq}
\end{figure}

The two previous limiting cases correspond to $\kappa\to 0$ and $\sigma\to 0$. For finite tension and bending rigidity one has, from equation (\ref{eq:cutoffsigmakappa}),
\begin{align}
    &\langle R_h(q) R_h(-q)\rangle_h = \frac{1}{8\pi^2\tilde{\sigma}^2} \left[ \int_{|l|<q_\sigma}\frac{d^2l}{(2\pi)^2}\frac{(q\times l)^4}{l^2(q-l)^2} + q_\sigma^4\int_{|l|>q_\sigma}\frac{d^2l}{(2\pi)^2}\frac{(q\times l)^4}{l^4(q-l)^4}\right]\\
    &=\frac{q_:^4q_\sigma^6}{16(2\pi)^3\tilde{\sigma}^2}\left[\frac{q_:^6+11q_:^4+37q_:^2+18}{q_:^2(q_:^2+4)^2}+q_:^2\log q_:-\frac{q_:^{10}+10q_:^8+30q_:^6+16q_:^4-32q_:^2+72}{q_:^3(q_:^2+4)^{5/2}}\log\left(\frac{q_:+\sqrt{q_:^2+4}}{2}\right)\right],\label{eq:fullRR}
\end{align}
where in the second line we adopted the convenient notation for the dimensionless momentum $q_:=q/q_\sigma$. In these units, the previous results are contained in the large and small momentum parts,
\begin{align}
    \langle R_h(q) R_h(-q)\rangle_h & \xrightarrow[q_:\to 0]{} \frac{3q_:^4q_\sigma^6}{16(2\pi)^3\tilde{\sigma}^2} = \frac{3q^4}{32(2\pi)^3}\left(\frac{2\sigma}{\kappa}\right)\tilde{\sigma}^{-2},\\
    & \xrightarrow[q_:\to \infty]{} \frac{9q_:^2q_\sigma^6}{32(2\pi)^3\tilde{\sigma}^2} = \frac{3q^2}{16(2\pi)^3}\left(\frac{3}{2}\right)\tilde{\kappa}^{-2},
\end{align}
where now the tension-dominated ($q_:\to 0$) regime is cutoff by $q_\sigma$, which appears in parenthesis. In the bending-dominated ($q_:\to\infty$) regime, the extra numerical factor arises from our choice of how to split the integral in (\ref{eq:cutoffsigmakappa}). These limits are also clear from figure \ref{fig:RRq}.

Let us evaluate the renormalized vortex-antivortex potential from the full one-loop correction (\ref{eq:fullRR}). The effective potential is given by
\begin{align}
    &\Gamma[\varphi;\rho_v]=\int\left[-\frac{1}{2}\varphi\left(\frac{\Delta}{\tilde{\gamma}}-\langle R_hR_h\rangle_h\right)\varphi+i\varphi\rho_v\right].
\end{align}
Minimizing over $\varphi$ for a vortex pair configuration $\rho_v=2\pi \delta(r-r_1)-2\pi \delta(r-r_2)$ gives the corrected potential
\begin{equation}
    \beta v(r) = \int d^2q e^{i\vec{q}\cdot \vec{r}}\left(-\frac{\tilde{\gamma}}{q^2}+\frac{(2\pi)^2\tilde{\gamma}^2}{q^4}\langle R_h(q)R_h(-q)\rangle_h\right),
\end{equation}
which is plotted in figure 2 of the main text. The potential has the logarithmic profile, with coefficient renormalized to $\gamma_\kappa$ on lengths large compared to $r_\sigma = 1/q_\sigma$.

To compute the average size of the vortex-antivortex pairs, we note that due to the large energy cost of these configurations, the expansion of the partition function as a vortex gas is dominated by the single-pair contribution
\begin{equation}
    Z\sim \int \frac{d^2x_1}{a^2}\frac{d^2x_2}{a^2}e^{-2\beta E_c-\beta v(|r_1-r_2|)}=\int \frac{d^2R}{a^2}\frac{d^2r}{a^2}e^{-2\beta E_c-\beta v(r)},
\end{equation}
where $R=\frac{r_1+r_2}{2}$, $r=r_2-r_1$. The RMS value of the vortex-antivortex distance is then given by
\begin{align}
    d^{RMS}=\sqrt{\langle r^2\rangle}=\sqrt{\frac{\int_a^\infty r^3 e^{-\beta v(r)}}{\int_a^\infty r^1 e^{-\beta v(r)}}}&\xrightarrow[r_\sigma\to 0]{}a\sqrt{1-(2-\pi\tilde{\gamma})^{-1}},\label{eq:RRMSflat}\\
    &\xrightarrow[r_\sigma\to \infty]{}a\sqrt{1-(2-\pi\tilde{\gamma}_\kappa)^{-1}},\label{eq:RRMSkappa}
\end{align}
with the intermediate regime $r_\sigma\sim a$ shown in figure 2. Note that for finite bending rigidity expression (\ref{eq:RRMSkappa}) is larger than (\ref{eq:RRMSflat}), reflecting the softening of the potential at short distances. This also means that (\ref{eq:RRMSkappa}) diverges at a lower temperature than the flat case $T_{BKT}=\frac{\pi}{2}\gamma$, indicating a lowering of the BKT transition temperature.

\newpage

\section{III. Sine-Gordon mapping and general topology}

The mapping to the Sine-Gordon theory consists on evaluating the sum over vortex configurations. Let the surface $M$ have general Euler characteristic $\chi(M)$. Then, as shown in the main text, the total charge of the relevant vortex configurations is $Q=\sum_n q_n=-\chi$. We separate out $Q$ charges and sum over neutral extra configurations,
\begin{align}
    Z&=\int\cD\phi e^{-\int d^2x\sqg\left[\frac{(\p\phi)^2}{2\gamma}+i R\phi\right]}\int\prod_{j=1}^Q \left(\frac{d^2x_j\sqg}{a^2}\right) e^{-EQ}e^{-i\sum_{j}^Q 2\pi\phi(x_j)}\bigg[1+\\
    &\hspace{8cm}+e^{-2E}\int \frac{d^2y_1\sqg}{a^2} e^{i2\pi\phi(y_1)}\int \frac{d^2y_2\sqg}{a^2} e^{-i2\pi\phi(y_2)}+...\bigg]\nonumber\\
    &=\int\cD\phi e^{-\int d^2x\sqg\left[\frac{(\p\phi)^2}{2\gamma}+i R\phi\right]}\int\prod_{j=1}^Q \left(\frac{d^2x_j\sqg}{a^2}\right) e^{-EQ}e^{-i\sum_{j}^Q 2\pi\phi(x_j)}\bigg[1+e^{-E}\int\frac{d^2y_1\sqg}{a^2}(e^{i2\pi\phi(y_1)}+e^{-i2\pi\phi(y_1)})+\nonumber\\
    &\hspace{5cm}+\1 e^{-2E}\int\frac{d^2y_1\sqg}{a^2}\frac{d^2y_2\sqg}{a^2}(e^{i2\pi\phi(y_1)}+e^{-i2\pi\phi(y_1)})(e^{i2\pi\phi(y_2)}+e^{-i2\pi\phi(y_2)})+...\bigg]\nonumber\\
    &=\int\cD\phi e^{-\int d^2x\sqg\left[\frac{(\p\phi)^2}{2\gamma}+i R\phi-\frac{2z}{a^2}\cos(2\pi\phi)\right]}\int\prod_{j=1}^Q \left(\frac{d^2x_j\sqg}{a^2}\right) e^{-EQ}e^{-i\sum_{j}^Q 2\pi\phi(x_j)},
\end{align}
where we defined the fugacity of vortices $z=e^{-E}$, and we rearranged the terms in the sum by making use of the charge screening sum rule. Let us define the averaging over the surface,
\begin{equation}
    f(\Bar{x}_1,...,\Bar{x}_N)=\int\prod_{j=1}^N\left(\frac{d^2x_j\sqg}{a^2}\right) f(x_1,...,x_N),
\end{equation}
and the $Q$-correlation function
\begin{equation}
    \langle\prod_{j=1}^Qe^{-i2\pi\phi(x_j)}\rangle_{\phi ;g}=\int\cD\phi\prod_{j=1}^Qe^{-i2\pi\phi(x_j)}e^{-S_{SG}[\phi,g]},
\end{equation}
where $S_{SG}$ is the Sine-Gordon action on a background metric $g$,
\begin{equation}
    S_{SG}[\phi,g]=\int d^2x\sqg\left[\frac{(\p\phi)^2}{2\gamma}+i R\phi-\frac{2z}{a^2}\cos(2\pi\phi)\right].\label{eq:sgphisupplementary}
\end{equation}
Then we find that $Z_\phi$ is given by 
\begin{equation}
    Z_{\phi}[g]=\left(\frac{z}{2}\right)^Q{\langle\prod_{j=1}^Qe^{-i2\pi\phi(\bar{x}_j)}\rangle_{\phi ;g}}.
\end{equation}
The limit of zero vortex fugacity $z\to 0$ is given by correlators of minimal models on Riemann surfaces, which have been studied for arbitrary genus \cite{riemannsurface}.

\newpage
\section{III. Renormalization group treatment}

We consider the effective potential
\begin{equation}
    \Gamma[\varphi,h] = W[\eta,\xi]-\int d^2x\varphi\eta-\int d^2x h\xi,\label{eq:FLegendre}
\end{equation}
evaluated from the full partition function
\begin{equation}
    Z=\int \cD\phi\cD h e^{-\int d^2x\left[\frac{(\p\phi)^2}{2\tilde{\gamma}}+i R_h\phi-\frac{2z}{a^2}\cos(2\pi\phi)+\frac{\tilde{\kappa}(\Delta h)^2}{2}+V_h\right]},
\end{equation}
where $W[\eta,\xi]=-\log\langle e^{\int d^2x(\phi\eta+h\xi)}\rangle$ is the generating functional. In the limit of a flat background $\kappa^{-1}\to 0$, the superfluid phase diagram is computed from the XY model RG flow equations around the free boson fixed point $\tilde{\gamma}=2/\pi,z=0$. Thus we consider the one-loop expansion of the effective potential around $(\kappa^{-1},\tilde{\gamma},z)=(0,2/\pi,0)$, to the combined second order in the vortex fugacity and in the amplitude of height fluctuations.

\subsection{$n$-point vertices}

The expansion of the effective potential in terms of the fields $(\chi^i)=(\phi,h)$ defines the $n$-point vertex $\mathbb{V}^{(n)}_{i_1,...,i_n}(x_1,...,x_n)$,
\begin{equation}
    \Gamma[\chi]=\sum_{n=2}^\infty\prod_{l=1}^n d^2x_l\mathbb{V}^{(n)}_{i_1,...,i_n}(x_1,...,x_n)\chi^{i_1}(x_1)...\chi^{i_n}(x_n).
\end{equation}
In particular, it follows from (\ref{eq:FLegendre}) that the $\phi-\phi$ two-point vertex is given by
\begin{align}
    &\mathbb{V}^{(2)}_{\phi,\phi}(x,x')=\frac{1}{\langle\phi(x)\phi(x')\rangle}, &\langle\phi(x)\phi(x')\rangle = -\frac{\delta^2W[\eta,\xi]}{\delta \eta(x)\delta \eta(x')}\Bigg|_{\eta=\xi=0}.\label{eq:vphiphi}
\end{align}
and the generating functional is
\begin{align}
    &e^{-W[\eta,0]}=\int\cD\phi\cD h\left[\sum_{n,p,q=0}^\infty\frac{1}{n!p!q!}\left(\frac{2z}{a^2}\int \cos 2\pi\phi\right)^n\left(-i\int R_h\phi\right)^p\left(-\int V_h\right)^q\right] e^{-S_0+\int \eta\phi},\label{eq:expandWick}
\end{align}
where $S_0$ is the free part of the action. It leads bare propagator term $n=p=q=0$,
\begin{equation}
    \langle\phi(x)\phi(x')\rangle_{(0,0,0)}=-\frac{\tilde{\gamma}}{\Delta}(x,x').\label{eq:phipropagator}
\end{equation}
From the $n=q=0$ terms, the $p=1$ term vanishes, and the $p=2$ term gives
\begin{align}
    \Big\langle\phi(x)\phi(x')\frac{1}{2}\left(i\int R_h\phi\right)^2\Big\rangle=&-\int d^2u \frac{\tilde{\gamma}}{(-\Delta)}(x,u) \int d^2u' \frac{\tilde{\gamma}}{(-\Delta)}(x',u') \langle R_h(u)R_h(u')\rangle .\label{eq:phi2RR}
\end{align}

We evaluate the curvature correlation functions in the nearly flat regime, where one can expand in the amplitude of height fluctuations, $\kappa^{-1}$. The bare $h$ propagator is
\begin{align}
    \cG(x,x')=\langle h(x)h(x')\rangle=\frac{1}{\tilde{\kappa}\Delta^2}(x,x')=\int\frac{d^2k}{(2\pi)^2}e^{ik\cdot(x-x')}\frac{1}{\tilde{\kappa} k^4},
\end{align}
and the one-point function of the curvature vanishes,
\begin{align}
    \langle R_h(x)\rangle_{h}=\frac{1}{2}[\Delta\Delta'\cG(x,x')-\p_{ij}\p_{ij}'\cG(x,x')]\Bigg|_{x=x'}=0,\label{eq:Rzero}
\end{align}
due to the translational invariance of the propagator $\cG(x,x')=\cG(x-x')$ in the nearly flat background. In other words, the average value of the curvature is zero as expected. On the other hand, the two-point function is given by
\begin{align}
    \langle R_h(x) R_h(x')\rangle_{h}=&\frac{1}{4}\langle[(\Delta h)^2-(\p^2_{ij}h)^2](x)[(\Delta h)^2-(\p^2_{ij}h)^2](x')\rangle_{h}\nonumber\\
    =&\frac{1}{2}[(\Delta\Delta'\cG(x,x'))^2-(\Delta\p_{ij}'\cG(x,x'))^2-(\p_{ij}\Delta'\cG(x,x'))^2+(\p_{ij}\p_{kl}'\cG(x,x'))^2]\nonumber\\
    =&\frac{1}{2}\int\frac{d^2l}{(2\pi)^2}\frac{d^2m}{(2\pi)^2}e^{i(l+m)\cdot(x-x')}[l^4m^4-2l^2m^2(l\cdot m)^2+(l\cdot m)^4]\cG(l)\cG(m)\nonumber\\
    =&\frac{1}{2}\int\frac{d^2l}{(2\pi)^2}\frac{d^2m}{(2\pi)^2}e^{i(l+m)\cdot(x-x')}[l^2m^2-(l\cdot m)^2]^2\cG(l)\cG(m)\nonumber\\
    =&\frac{1}{2}\int\frac{d^2l}{(2\pi)^2}\frac{d^2m}{(2\pi)^2}e^{i(l+m)\cdot(x-x')}(l\times m)^4\cG(l)\cG(m).\label{eq:RR}
\end{align}
Substituting back in (\ref{eq:phi2RR}), we find
\begin{align}
    \Big\langle\phi(x)\phi(x')\frac{1}{2}\left(i\int \frac{R_h\phi}{2\pi}\right)^2\Big\rangle &=-\int\frac{d^2q}{(2\pi)^2}e^{iq\cdot(x-x')}\left(\frac{\tilde{\gamma}}{q^2}\right)^2\int\frac{d^2l}{(2\pi)^2}\frac{(l\times q)^4\cG(l)\cG(q-l)}{2}\nonumber\\
    &=-\int\frac{d^2q}{(2\pi)^2}e^{iq\cdot(x-x')}\left(\frac{\tilde{\gamma}}{q^2}\right)^2\frac{1}{2\tilde{\kappa}^2}\int\frac{d^2l}{(2\pi)^2}\frac{(l\times q)^4}{l^4(q-l)^4}.\label{eq:phicorrelation1}
\end{align}
Using Feynman parameters, the remaining loop integral evaluates to
\begin{align}
    \int d^2l\frac{(l\times q)^4}{l^4(q-l)^4}&=\int d^2l (l\times q)^4\frac{\Gamma(4)}{\Gamma(2)\Gamma(2)}\int_0^1 dx dy \frac{\delta(1-x-y)xy}{[xl^2+y(q-l)^2]^4}=\frac{3\pi}{4}q^2.\label{eq:momentumint}
\end{align}
Thus the $p=2$ term of (\ref{eq:vphiphi}) is
\begin{align}
    \Big\langle\phi(x)\phi(x')\frac{1}{2}\left(i\int R_h\phi\right)^2\Big\rangle=-\int\frac{d^2q}{(2\pi)^2}e^{iq\cdot(x-x')}\left(\frac{\tilde{\gamma}}{q^2}\right)^2\frac{3q^2}{32\pi\tilde{\kappa}^2}.\label{eq:phipropcorrection}
\end{align}
Note that (\ref{eq:phipropcorrection}) is suppressed by a factor of $\kappa^{-2}$ with respect to the bare propagator (\ref{eq:phipropagator}). This the perturbative meaning of the expansion (\ref{eq:expandWick}): although the $R_h\phi$ and $V_h$ terms do not have small coupling constants, the correlation functions involving higher powers of $R_h$ are suppressed by the higher powers of $\kappa^{-1}$ which accompany the $h$ propagators. To this order, the $\phi-\phi$ vertex is given by
\begin{align}
    \mathbb{V}^{(2)}_{\phi\phi}(q)=& \frac{q^2}{\tilde{\gamma}}\left[1-\frac{\tilde{\gamma}}{q^2}\frac{3\pi q^2}{32\pi\tilde{\kappa}^2}\right]^{-1}=\frac{q^2}{\tilde{\gamma}}+\frac{3\pi q^2}{32\pi\tilde{\kappa}^2}+O(\kappa^{-4})=\frac{q^2}{\tilde{\gamma}_\kappa}+O(\kappa^{-4}),\label{eq:vphiphifull}
\end{align}
where
\begin{equation}
    \gamma_\kappa^{-1} = \gamma^{-1} + \frac{3T}{32\pi\kappa^2}.\label{eq:gammacorrection}
\end{equation}
As noted in the main text, the effective potential looks like the bare one but with a finite shift of the superfluid stiffness, which is lowered by the averaging over shape fluctuations of the background corresponding to the diagram in figure 3 a). This correction always appears whenever the $\phi$ propagator figures as a sub-diagram of a given amplitude. Note also that the corrections to the $\phi-\phi$ vertex due to the higher-order terms of the bending energy, $V_h$, appear only at two-loop order $O(\kappa^{-4})$.

At first order in $z$, one finds
\begin{align}
    \Big\langle\phi(x)\phi(x')\left(\frac{2z}{a^2}\int \cos 2\pi\phi\right)\Big\rangle=&\frac{2z}{a^2}\sum_{n=0}^\infty \frac{(-1)^n}{(2n)!}(2\pi)^{2n}\int d^2u \langle\phi(x)\phi(x')\phi^{2n}(u)\rangle\nonumber\\
    =&\frac{2z}{a^2}\sum_{n=1}^\infty\frac{(-1)^n}{(2n)!}\int d^2u \frac{(2n)!(2\pi)^{2n}}{(n-1)!2^{n-1}}\langle\phi(x)\phi(u)\rangle_0\langle\phi(u)\phi(x')\rangle_0\langle\phi(u)\phi(u)\rangle_0^{n-1}\nonumber\\
    =&-\frac{2z}{a^2}\int d^2u \langle\phi(x)\phi(u)\rangle_0\langle\phi(u)\phi(x')\rangle_0 (2\pi)^2 e^{-2\pi^2G(0)},\label{eq:z1y0divergence}
\end{align}
where we dropped the disconnected terms and included in the second line the symmetry factors corresponding to different ways to get the same contraction. These are the tadpole contributions of the XY model, which involve the UV divergent closed loops with factors of $G^\phi(0)=\langle\phi(u)\phi(u)\rangle$ \cite{coleman75sinegordon,amit1980renormalisation,wiegmann1978one}. The dependence on the cutoff scale $a$ is
\begin{align}
    G^\phi(0)=\tilde{\gamma}\lim_{|x|\to a}\int\frac{d^2q}{(2\pi)^2}\frac{e^{iq\cdot x}}{q^2+\mu^2}=\tilde{\gamma}\frac{1}{2\pi}K_0(\mu a)\approx -\frac{\tilde{\gamma}}{2\pi}\ln\left(\frac{e^{\gamma_E}}{2}\mu a\right),\label{eq:gphi0regularization}
\end{align}
where we introduced a mass scale $\mu$ as an IR regulator and $\gamma_E\approx 0.58$ is Euler's constant. Thus (\ref{eq:z1y0divergence}) diverges as
\begin{align}
    &\frac{2z}{a^2}\left(\frac{e^{\gamma_E}}{2}\mu a\right)^{\pi\tilde{\gamma}}=2z\zeta, &\zeta = c\mu^2(c\mu^2a^2)^{\frac{\pi}{2}\tilde{\gamma}-1},\label{eq:zeta}
\end{align}
with $c=e^{2\gamma_E}/4$. With these definitions, (\ref{eq:z1y0divergence}) becomes
\begin{align}
    \Big\langle\phi(x)\phi(x')\left(\frac{2z}{a^2}\int \cos 2\pi\phi\right)\Big\rangle=-2z\zeta\int\frac{d^2q}{(2\pi)^2}e^{iq\cdot(x-x')}\left[\frac{\tilde{\gamma}}{q^2}\right]^2.\label{eq:phiphiorderz}
\end{align}
The leading correction to this contribution coming from fluctuations of the height field is
\begin{align}
    \Big\langle\phi(x)\phi(x')\left(\frac{2z}{a^2}\int \cos 2\pi\phi\right)\frac{1}{2}\left(-i\int R_h\phi\right)^2\Big\rangle,
\end{align}
which changes the expansion (\ref{eq:z1y0divergence}) by modifying either of the $\phi$ propagators as in (\ref{eq:phipropcorrection}). Thus, up to $O(\kappa^{-2})$ corrections, we again have the substitution (\ref{eq:gammacorrection}). The correction (\ref{eq:phiphiorderz}) gives the $O(z)$ term of (\ref{eq:vphiphifull}),
\begin{align}
    \mathbb{V}^{(2)}_{\phi\phi}(q) \to \frac{q^2}{\tilde{\gamma}_\kappa}+2z\zeta\label{eq:vz}.
\end{align}

At the next order in the vortex fugacity, again insertions at the same point can be contracted, generating powers of $G^\phi(0)=\langle\phi(u)\phi(u)\rangle$, the tadpole diagrams. Factoring out all tadpoles renormalizes $z$ according to (\ref{eq:zeta}), so that the remaining diverging connected contractions are given by
\begin{align}
    &\frac{1}{2!}\left(\frac{2z}{a^2}\right)^2\Big\langle\int d^2u d^2u'\cos\phi(u)\cos(u')\phi(x)\phi(x')\Big\rangle=\nonumber\\
    &\hspace{2cm}= \frac{(2z\zeta)^2}{2}\int d^2u d^2u'\Big[\langle\phi(x)\phi(u)\rangle_0\langle\phi(u')\phi(x')\rangle_0\sum_{n=1}^\infty\frac{\langle\phi(u)\phi(u')\rangle_0^{2n+1}}{(2n+1)!}\nonumber\\
    &\hspace{6.5cm}+\langle\phi(x)\phi(u)\rangle_0\langle\phi(u)\phi(x')\rangle_0\sum_{n=1}^\infty\frac{\langle\phi(u)\phi(u')\rangle_0^{2n}}{(2n)!}\Big].
\end{align}
The terms with odd powers of $\langle\phi(u)\phi(u')\rangle_0$ give
\begin{align}
    2(z\zeta)^2\int\frac{d^2q}{(2\pi)^2}e^{iq\cdot(x-x')}\left[\frac{\tilde{\gamma}}{q^2}\right]^2[\sinh G^\phi-G^\phi](q),
\end{align}
while the terms with even powers of $\langle\phi(u)\phi(u')\rangle_0$ give
\begin{align}
    -2(z\zeta)^2\int\frac{d^2q}{(2\pi)^2}e^{iq\cdot(x-x')}\left[\frac{\tilde{\gamma}}{q^2}\right]^2[\cosh G^\phi-1](0).
\end{align}
Thus to second order in vortex fugacity and in the inverse rigidity of the background, (\ref{eq:vphiphi}) is
\begin{align}
    \mathbb{V}^{(2)}_{\phi\phi}(q)=\frac{q^2}{\tilde{\gamma}_\kappa}+2z\zeta+2(z\zeta)^2\{[\cosh G^\phi-1](0)-[\sinh G^\phi-G^\phi](q)\}.\label{eq:vphiresult}
\end{align}

\subsection{Regularization and RG equations}

The $\phi-\phi$ two-point vertex (\ref{eq:vphiresult}) diverges in the UV limit $a\to 0$. For example, from (\ref{eq:zeta}),
\begin{align}
    \zeta=c\mu^2(c\mu^2a^2)^\delta=c\mu^2[1+\delta\log(c\mu^2a^2)+O(\delta^2)],\label{eq:zetaexp}
\end{align}
where $\delta=\frac{\pi}{2}\tilde{\gamma}_\kappa-1$ is the deviation from the free boson fixed point $\tilde{\gamma}_\kappa=2/\pi$. We regularize the UV divergences to combined second order in the small $z,\delta$ parameters. Besides the divergence due to the $\zeta$ parameter, (\ref{eq:vphiresult}) contains the diverging integral
\begin{align}
    \int d^2x x^2[\sinh G^\phi(x)-G^\phi(x)]&=\int_0^{\mu^{-1}}\frac{\pi x^3 dx}{2[c\mu^2(x^2+a^2)]^{2(1+\delta)}}=\frac{\pi}{4cm\phi^2}\int_0^{\mu^{-2}}\frac{wdw}{(w+a^2)^2}\nonumber\\
    &=-\frac{\pi\log(c\mu^2a^2)}{4c\mu^2},\label{eq:sinhlog}
\end{align}
which appears when expanding the Fourier transform in (\ref{eq:vphiresult}). Here, we dropped regular terms and used $\delta\approx 1$ since this term appears multiplying $z^2$. Thus we find the divergent part of $\mathbb{V}^{(2)}_{\phi\phi}(q)$,
\begin{align}
    \mathbb{V}^{(2)}_{\phi\phi}(q)=\frac{q^2}{\tilde{\gamma}_\kappa}+2zc\mu^2[1+\delta\log(c\mu^2a^2)]-\frac{\pi z^2}{2}q^2\log(c\mu^2a^2).
\end{align}
We proceed by defining renormalized constants $\tilde{\gamma}_{\kappa,r},z_r$ by $\tilde{\gamma}_\kappa=Z_\gamma\tilde{\gamma}_{\kappa,r}$, $z=Z_z z_r$ and absorbing the divergences into the definitions of $Z_\gamma,Z_z$ to leading order in $\delta_r,z_r$, which gives
\begin{align}
    Z_\gamma(\mu)=& 1+2\pi^3 \tilde{\gamma}_{\kappa,r} z_r^2\log(\mu^2a^2),\label{eq:Zgamma}\\
    Z_z(\mu)=& 1-\left(\frac{\pi}{2}\tilde{\gamma}_{\kappa,r}-1\right)\log(\mu^2a^2).\label{eq:Zmu}
\end{align}
Finally, under the renormalization transformation $\mu=\frac{e^{-l}}{a}$ one finds the flow equations of the renormalized constants
\begin{align}
    \frac{d\tilde{\gamma}_{\kappa,r}^{-1}}{dl}=&\frac{\tilde{\gamma}_{\kappa}}{Z_\gamma^2}\frac{dZ_\gamma}{d\log\mu}=\frac{4\pi^3}{Z_\gamma}z_r^2,\\
    \frac{dz_r}{dl}=&\frac{z}{Z_z^2}\frac{dZ_z}{d\log\mu}=\frac{2z_r}{Z_z}\left(1-\frac{\pi}{2}\tilde{\gamma}_{\kappa,r}\right).
\end{align}
Evaluating at the bare parameters gives the first pair of RG equations,
\begin{align}
    \frac{d\tilde{\gamma}_{\kappa}}{dl}&=-4\pi^3\tilde{\gamma}_{\kappa}^2z^2,\label{eq:BKT1S}\\
    \frac{dz}{dl}&=\left(2-\pi\tilde{\gamma}_{\kappa,r}\right)z.\label{eq:BKT2S}
\end{align}

\subsection{Renormalization of the bending rigidity}

We can also use the expansion (\ref{eq:expandWick}) to evaluate the $\langle h(x)h(x')\rangle$ correlation function. In this case, the correction to the bending energy $V_h$ plays a role at one loop,
\begin{align}
    -\int d^2u&\langle h(x)h(x') V_h(u)\rangle=-\frac{\kappa}{4}\int d^2u\langle h(x)h(x') (4\Delta h \bp^\mu h\p^2_{\mu\nu}h\bp^\nu h-5(\Delta h)^2(\p h)^2)(u)\rangle,
\end{align}
which corresponds to the diagram in figure 3 b) of the main text. One can expand all contractions as in (\ref{eq:RR}). These are of two different types, which give corrections proportional to $q^2$ and to $q^4$ to the $h-h$ two-point vertex. The first type renormalizes the tension, which we then set to zero. The second type renormalizes the bending stiffness and is given by
\begin{align}
    \int\frac{d^2q}{(2\pi)^2}\frac{e^{iq\cdot(x-x')}}{(\kappa q^4)^2}\int\frac{d^2l}{(2\pi)^2}\frac{\{5q^4l^2-4q^2(q\times l)^2\}}{2l^4}=-\int\frac{d^2q}{(2\pi)^2}\frac{e^{iq\cdot(x-x')}}{(\kappa q^4)^2}\frac{3q^4}{8\pi}\log(c\mu^2 a^2),\label{eq:hh100}
\end{align}
which is of order $\kappa^{-2}$. Here, we used the fact that $\bar{q}^\mu l_\mu=\epsilon^{\mu\lambda}q_\lambda l_\nu=q\times l$ and regularized the remaining logarithmically divergent integral in the IR by the scale $\mu$ and in the UV by the short-distance cutoff $a$ as in (\ref{eq:gphi0regularization}).

On the other hand, the contribution shown in figure 3 c) evaluates to
\begin{align}
    \Big\langle h(x)h(x')\left[\frac{1}{2!}\left(-i\int R_h\phi\right)^2\right]\Big\rangle&=-\int\frac{d^2q}{(2\pi)^2}\frac{e^{iq\cdot(x-x')}}{(\tilde{\kappa} q^4)^2}\frac{\gamma}{\kappa}\int\frac{d^2l}{(2\pi)^2}\frac{(q\times l)^4}{l^2(q-l)^4}\label{eq:hhRR}\\
    &=\int\frac{d^2q}{(2\pi)^2}\frac{e^{iq\cdot(x-x')}}{(\tilde{\kappa} q^4)^2}\frac{3q^4\gamma}{32\pi\kappa}\log(c\mu^2 a^2),\label{eq:hh020}
\end{align}
Finally, higher powers in $z$ can be computed from the corresponding corrections to the $\phi-\phi$ vertex, for example,
\begin{align}
    &\Big\langle h(x)h(x')\left(\frac{2z}{a^2}\int \cos 2\pi\phi\right)\frac{1}{2!}\left(-i\int R_h\phi\right)^2\Big\rangle=\nonumber\\
    &\hspace{3cm}=-\frac{1}{2}\int d^2u d^2u'\Big\langle\phi(u)\left(\frac{2z}{a^2}\int \cos 2\pi\phi\right)\phi(u')\Big\rangle\langle h(x)h(x')R_h(u)R_h(u')\rangle\nonumber\\
    &\hspace{3cm}=-\frac{1}{2}\int d^2u d^2u'\left(-2z\zeta\int\frac{d^2l}{(2\pi)^2}e^{il\cdot(x-x')}\left[\frac{\tilde{\gamma}}{l^2}\right]^2\right)\langle h(x)h(x')R_h(u)R_h(u')\rangle\nonumber\\
    &\hspace{3cm}=2z\tilde{\gamma}^2\zeta\int\frac{d^2q}{(2\pi)^2}\frac{e^{iq\cdot(x-x')}}{(\tilde{\kappa} q^4)^2}\frac{3\pi}{2\tilde{\kappa}}q^4,
\end{align}
and similarly for the $z^2$ contribution. Finally, one has the $h$-$h$ two-point vertex
\begin{align}
    \mathbb{V}^{(2)}_{hh}(q)=&\tilde{\kappa} q^4+\frac{3q^4}{8\pi}\log(c\mu^2 a^2)-\frac{3\gamma q^4}{32\pi\kappa}\log(c\mu^2 a^2)-\frac{3z\zeta\tilde{\gamma}^2\pi q^4}{\tilde{\kappa}}+\\
    &\hspace{3cm}+\frac{2(z\zeta\tilde{\gamma})^2}{\tilde{\kappa}}\int d^2l \frac{(q\times l)^4\{[\sinh G^\vartheta-G^\vartheta](l)-[\cosh G^\vartheta-1](0)\}}{l^4(q-l)^4}.\label{eq:v2hhfull}
\end{align}
Through a similar procedure as for the $\phi-\phi$ vertex, one can absorb the logarithmic diverge of this expression in terms of a renormalization of the bending rigidity $\kappa=Z_\kappa\kappa_r$ as
\begin{align}
    Z_\kappa(\mu)=1-\frac{3}{8\pi\kappa_r}\log(\mu^2a^2)+\frac{3\tilde{\gamma}_{\kappa,r}}{32\pi\tilde{\kappa}^2_r}\log(\mu^2a^2)+\frac{3\pi^2z_r^2\tilde{\gamma}_{\kappa,r}^2}{32\kappa_r^2}(\log(\mu^2a^2))^2.
\end{align}
Then the flow with the scale $\mu=\frac{e^{-l}}{a}$, expanded around the bare values, leads to the RG flow equation for $\kappa$,
\begin{equation}
    \frac{d\kappa}{dl}=-\frac{3T}{4\pi}\left(1-\frac{\gamma_{\kappa}}{4\kappa}\right).\label{eq:kappaflowS}
\end{equation}

\pagebreak
\section{Crystalline and hexatic order}

Note that, in the abscence of the chiral superfluid, the equation (\ref{eq:kappaflow}) becomes
\begin{equation}
    \frac{d\kappa}{dl} = -\frac{3T}{4\pi},
\end{equation}
which is characteristic of fluid membranes. It predicts that the bending rigidity flows to zero at large scales. Integrating this equation from $\kappa_0$ at the scale $a$ to $0$ at scale $\xi$ where it vanishes, we find that the finite correlation length is given by
\begin{equation}
    \xi = a e^{4\pi\kappa_0/3T}.
\end{equation}
It is well-known that the presence of in-plane order can change the RG flow leading to a stable flat phase at low temperatures. An example is in-plane crystalline order, for which the modification of the RG equations can be understood as follows. In linear elasticity theory, the elastic energy leads to the Boltzmann factor
\begin{equation}
    e^{-\beta E_u}=e^{-\frac{1}{2}\int d^2x \tilde{C}_{ijkl}u_{ij}u_{kl}}=\int \cD\sigma e^{-\int d^2x\left[ \frac{1}{2}\tilde{C}^{-1}_{ijkl}\sigma_{ij}\sigma_{kl}+i\sigma_{ij}u_{ij}\right]},
\end{equation}
where $u_{ij}$ is the strain tensor, $\tilde{C}_{ijkl}$ is the elasticity tensor in units of temperature $k_BT$, and in the second step we introduced the stress tensor $\sigma_{ij}$ as a Hubbard-Stratonovich dual. The embedding map is defined with respect to deviations $u_i(x)$ from the equilibrium positions of the two-dimensional lattice as $\mathbf{R} = (x_i+u_i(x),h(x))$ so that, for small in-plane deformations,
\begin{equation}
    u_{ij} = \frac{1}{2}(\p_i\mathbf{R}\cdot\p_j\mathbf{R}-\delta_{ij})\approx \frac{1}{2}(\p_i u_j+\p_j u_i+\p_i h\p_j h).\label{eq:strainuh}
\end{equation}
Before integrating out phonons, it is important to separate the parallel and transverse parts of the symmetric tensor $\p_ih\p_jh$. Consider a two-dimensional symmetric tensor field
\begin{equation}
    \eta_{ij}(x)=\eta_{ji}(x),\label{eq:symeta}
\end{equation}
and let us denote by $\eta^F_{ij}(p)$ its Fourier transform. In two dimensions, one can show that it has a decomposition
\begin{equation}
    \eta^F_{ij}(p)=p_i\xi_j+p_j\xi_i+\bar{p}_i\bar{p}_j\delta,
\end{equation}
for some $\xi_i$, $\delta$. The corresponding decomposition of $\p_ih\p_jh$ in coordinate space is
\begin{equation}
    \p_ih\p_jh=\p_iv_j+\p_jv_i+P^T_{ij}f,\label{eq:decomposedh}
\end{equation}
with the transverse projector
\begin{equation}
    P^T_{ij}=\frac{\bar{\p}_i\bar{\p}_j}{\Delta}=\delta_{ij}-\frac{\p_i\p_j}{\Delta}.
\end{equation}
Substituting the decomposition (\ref{eq:decomposedh}) into the expression for the strain field (\ref{eq:strainuh}), we see that the longitudinal part $v_i$ can be absorbed into a shift of the phonon field $u_i^p$. The transverse part,
\begin{equation}
    f=P^T_{ij}(\p_i h\p_j h)=\Delta^{-1}[(\p^2_{ij}h)^2-(\Delta h)^2]=-\frac{2}{\Delta}R_h,\label{eq:RhPT}
\end{equation}
is related to the Gaussian curvature $R_h$.

Integrating out the phonon fluctuations $u_i^p$ gives
\begin{equation}
    \p_i\sigma_{ij}=0, \,\,\, \sigma_{ij}=\sigma_{ji},
\end{equation}
so that, in the absence of external stress, the longitudinal component of $\sigma_{ij}$ vanishes. The remaining transverse component,
\begin{equation}
    \sigma_{ij}=\bar{\p}_i\bar{\p}_j\chi,
\end{equation}
defines the Airy stress potential $\chi$. The remaining integral is
\begin{align}
    \int\cD\chi e^{-\int d^2x\left[\frac{1}{2}\tilde{C}^{-1}_{ijkl}\bar{\p}_i\bar{\p}_j\chi\bar{\p}_k\bar{\p}_l\chi+i\chi\bar{\p}_i\bar{\p}_j(\frac{1}{2}\p_ih\p_jh)\right]}=\int\cD\chi e^{-\int d^2x\left[\frac{1}{2\tilde{Y}}(\Delta\chi)^2-i\chi R_h\right]},
\end{align}
where we considered the case of a hexagonal lattice, for which the elastic tensor has the isotropic form
\begin{equation}
    C_{ijkl}=2\mu\delta_{ik}\delta_{jl}+\lambda\delta_{ij}\delta_{kl},
\end{equation}
and $Y$ denotes the two-dimensional Young modulus
\begin{equation}
    Y=\frac{4\mu(\mu+\lambda)}{2\mu+\lambda},\label{eq:youngY}
\end{equation}
with $\tilde{Y}=\beta Y$. Finally, integrating out the Airy stress potential leads to a long-range interaction of the out-of-plane fluctuations,
\begin{equation}
    \frac{\tilde{Y}}{2}\int\frac{d^2q}{(2\pi)^2}\frac{R_h(q)R_h(-q)}{q^4}.\label{eq:YRhRh}
\end{equation}
In the case of a hexatic membrane, crystalline order is melted out but orientational order remains, and one can derive a similar interaction,
\begin{equation}
    \frac{\tilde{J}}{2}\int\frac{d^2q}{(2\pi)^2}\frac{R_h(q)R_h(-q)}{q^2},\label{eq:JRhRh}
\end{equation}
where $J$ denotes the hexatic stiffness. Note that the effective interaction is of the same form but less long-ranged than that stemming from crystalline order.
\begin{figure}[h!]
    \centering 
    \includegraphics[width=\textwidth]{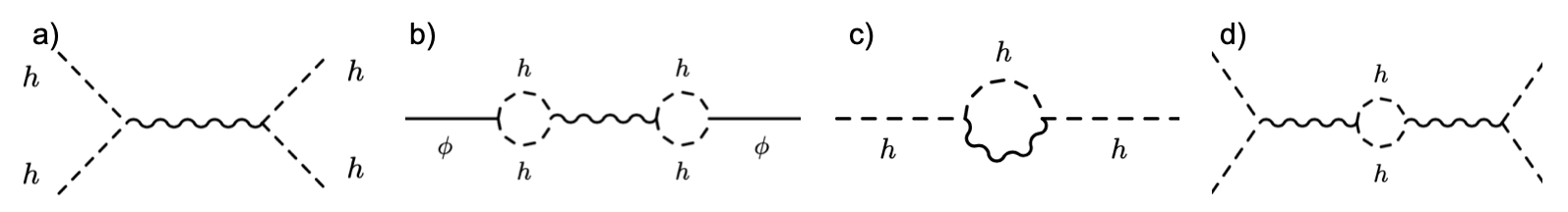}
    \caption{a) Interaction vertex arising from in-plane (crystalline or hexatic) order. b) Lowest-order correction to the $\phi-\phi$ vertex appears only at two loops. c) Renormalization of the membrane rigidity by the in-plane order vertex. d) One-loop renormalization of the in-plane order vertex.}
    \label{fig:inplane}
\end{figure}
Let us represent the effective vertex due to in-plane order (\ref{eq:YRhRh}), (\ref{eq:JRhRh}) as in figure \ref{fig:inplane} a). Then the lowest-order correction to the $\phi-\phi$ vertex due to the in-plane interaction appears only at two-loop order $\kappa^{-4}$, as shown in figure \ref{fig:inplane} b). Thus to leading order the modified RG equations for the BKT transition of the chiral superfluid, (\ref{eq:BKT1}), (\ref{eq:BKT2}), are not modified. On the other hand, the RG flow equation for the bending rigidity is modified by the one-loop correction to the $h-h$ vertex in figure \ref{fig:inplane} c). Finally, the in-plane interaction vertex is also modified at one loop by the contribution in figure \ref{fig:inplane} d).

In the hexatic case, the effective interaction (\ref{eq:JRhRh}) is equivalent to the one mediated by the $\phi$ field, so that the effect of the extra in-plane order can be easily understood. The correction to the $h-h$ vertex is equivalent to the one evaluated from (\ref{eq:phicorrelation1}), and leads to an extra renormalization of the bending rigidity
\begin{equation}
    \frac{d\kappa}{dl}=-\frac{3T}{4\pi}\left(1-\frac{\gamma_{\kappa}}{4\kappa}-\frac{J_{\kappa}}{4\kappa}\right),
\end{equation}
while the diagram in figure \ref{fig:inplane} d) leads to a finite renormalization $J_\kappa^{-1} = J^{-1} + \frac{3T}{32\pi\kappa^2}$ analogous to (\ref{eq:gammacorrection}). This modification allows one to probe the BKT transition of the superfluid into a phase where, although the proliferation of vortices has destroyed superfluidity, the background is still flat. And in this context, one can ask how the superfluidity changes the properties of the flat phase. For the case of crystalline order, the modification of the flow equation for the bending rigidity also leads to a flat phase in the absence of superfluidity. However, in order to probe the changes in the flat phase due to the chiral superfluid, the renormalization procedure should be changed to account for the different critical dimension of crystalline membranes. The difference can be seen from the longer-ranged potential (\ref{eq:YRhRh}) as compared to (\ref{eq:JRhRh}). This is an interesting question for future investigation.


\end{document}